\documentclass[aps,preprint,prd,showpacs,nofootinbib]{revtex4}

\usepackage{amsmath,amssymb,amsfonts,array}
\usepackage{bm,bbm}
\usepackage{cases}
\usepackage{dcolumn}
\usepackage{graphicx}
\usepackage{hyperref}
\usepackage{subfigure}
\usepackage{wasysym}
\usepackage{xcolor}
\usepackage{ulem}

\hypersetup{colorlinks=true,
    breaklinks=true,
    pdfstartview=Fit,
    linkcolor=blue,
    citecolor=blue,
    urlcolor=blue}

\def\be{\begin{equation}}
\def\ee{\end{equation}}
\def\ba{\begin{eqnarray}}
\def\ea{\end{eqnarray}}

\def\blue{\color{blue}}

\def\nn{\nonumber}
\def\lf{\left}
\def\rt{\right}

\begin{document}

\title{Intermittent null energy condition violations during inflation and primordial gravitational waves}

\author{Yong Cai$^{1}$\footnote{\texttt{\blue yongcai\_phy@outlook.com}}}
\author{Yun-Song Piao$^{2,3,4,5}$\footnote{\texttt{\blue yspiao@ucas.ac.cn}}}

\affiliation{$^1$ School of Physics and Microelectronics, Zhengzhou University, Zhengzhou, Henan 450001, China}

\affiliation{$^2$ School of Physics, University of Chinese Academy
of Sciences, Beijing 100049, China}

\affiliation{$^3$ School of Fundamental Physics and Mathematical
Sciences, Hangzhou Institute for Advanced Study, UCAS, Hangzhou
310024, China}

\affiliation{$^4$ International Center for Theoretical Physics
Asia-Pacific, Beijing/Hangzhou, China}

\affiliation{$^5$ Institute of Theoretical Physics, Chinese
Academy of Sciences, P.O. Box 2735, Beijing 100190, China}

\begin{abstract}

Primordial null energy condition (NEC) violation
would imprint a blue-tilted spectrum on gravitational wave
background (GWB). However, its implications on the GWB
might be far richer than expected. We present a scenario, in which
after a slow-roll (NEC-preserving) inflation with Hubble parameter
$H\simeq H_{inf1}$, the Universe goes through an NEC-violating
period and then enters subsequent slow-roll inflation with a
higher $H$ ($=H_{inf2}\gg H_{inf1}$). The resulting primordial
gravitational wave spectrum is nearly flat at the cosmic microwave
background band, as well as at the frequency $f\sim 1/{\rm
yr}$ but with higher amplitude (compatible with the recent
NANOGrav result). It is also highlighted that for the multi-stage
inflation if the NEC violations happened intermittently, we
might have a Great Wall-like spectrum of the stochastic GWB at the
corresponding frequency band.

\end{abstract}

\maketitle
\tableofcontents

\section{Introduction}

The primordial gravitational wave background (GWB) \cite{Starobinsky:1979ty,Rubakov:1982df} with a broad frequency-band ($10^{-18}-10^{10}$ Hz) carries rich
information about the early Universe. It is usually thought that
its detection will not only solidify our confidence in inflation,
but also offer us an unparalleled probe to the physics related to the cosmological
(non)singularity, in which the null energy condition (NEC) violation might play
a significant role
\cite{Rubakov:2014jja,Libanov:2016kfc,Kobayashi:2016xpl,Ijjas:2016vtq,Dobre:2017pnt,Cai:2016thi,Creminelli:2016zwa,Cai:2017tku,Cai:2017dyi,Kolevatov:2017voe,Ye:2019sth},
and the UV-complete gravity theory.

The primordial gravitational waves (GWs) at the ultra-low
frequency band ($10^{-18}-10^{-16}$ Hz) would induce the B-mode
polarization in the cosmic microwave background (CMB). The search
for the primordial GWs with CMB has been still in progress. The
Pulsar Timing Array (PTA) experiments focus on GWB at frequencies
$f\sim 1/{\rm yr}$ ($\sim 10^{-8}$ Hz). Recently, based on the
$12.5$-yr data analysis, the NANOGrav Collaboration reported
evidence for a stochastic \textit{common-spectrum} process
\cite{Arzoumanian:2020vkk}, which might be interpreted as a
stochastic GWB with a spectrum tilt $-1.5\lesssim n_T\lesssim
0.5$, see
\cite{Vagnozzi:2020gtf,Li:2020cjj,Tahara:2020fmn,Kuroyanagi:2020sfw}
for the implications of NANOGrav's result in inflation. The
current bound on GWB at CMB band indicates a tensor-to-scalar
ratio $r\lesssim 0.06$ \cite{Ade:2018gkx}. Therefore, only if the
primordial GWs have a blue-tilted spectrum, it is able to be
detected by the experiments and detectors at other frequency
bands.

It is well-known that for inflation, if initially the GW modes sit
in the Bunch-Davis state (or e.g.,
\cite{Li:2020cjj,Ashoorioon:2014nta}), a blue-tilted spectrum
suggests that the corresponding inflation is inevitably
NEC-violating, i.e., $T_{\mu\nu}n^{\mu}n^{\nu}<0$, which
corresponds to ${\dot H}>0$, namely, super-inflation\footnote{In this paper, the term super-inflation is used for the accelerated expansion in which $\epsilon=-\dot{H}/H^2<0$, since $\ddot{a}/a=H^2(1-\epsilon)$.}
\cite{Piao:2003ty,Piao:2004tq,Baldi:2005gk,Piao:2006jz}.
The NEC-violating inflation may be performed stably with the Galileon
theory \cite{Kobayashi:2010cm,Kobayashi:2011nu} and the effective
field theory (EFT) of inflation
\cite{Creminelli:2006xe,Capurri:2020qgz}.
If initially the
NEC is violated drastically (${\dot H}\gg H^2$), it is also
possible that our Universe is asymptotically Minkowskian and
slowly expanding in infinite past \cite{Piao:2003ty}. In such
scenarios, the hot ``big bang'' evolution or inflation starts
after the end of the slow expansion or Genesis
\cite{Creminelli:2010ba,Liu:2011ns,Wang:2012bq,Liu:2012ww,Creminelli:2012my,Hinterbichler:2012fr,Hinterbichler:2012yn,Liu:2014tda,Pirtskhalava:2014esa,Nishi:2015pta,Kobayashi:2015gga,Cai:2016gjd,Nishi:2016ljg,Mironov:2019qjt,Ageeva:2020gti,Ilyas:2020zcb}.
Based on the beyond-Horndeski EFT (see e.g.
\cite{Langlois:2018dxi,Kobayashi:2019hrl} for reviews), the
Genesis could be implemented without pathologies (including
instabilities and superluminality)
\cite{Cai:2016thi,Creminelli:2016zwa,Cai:2017tku,Cai:2017dyi,Kolevatov:2017voe,Ye:2019sth}.

Inspired by current (and upcoming) experiments searching for GWB,
it is significant to resurvey the imprints of NEC violation on
stochastic GWB. Recently, a scenario in which the super-inflation
is followed by a slow-roll (NEC-preserving) inflation has been
proposed in \cite{Tahara:2020fmn}, which yields a large stochastic
GWB with $n_T\simeq 0.9$ at the PTA band. But the spectrum of
scalar perturbations, i.e., $P_s$, is highly blue-tilted too, since $n_s-1\simeq
n_T$. Consequently, other fields must be responsible for the density
perturbation at the CMB band. However, it is possible that a
slow-roll (NEC-preserving) inflation with $H=H_{inf1}$, which
results in $P_s\sim H^2_{inf1}/\epsilon\sim 10^{-9}$ at the CMB band,
happened before the NEC-violating phase, which is subsequently
followed by a slow-roll inflation with a higher scale $H_{inf2}\gg
H_{inf1}$. In Refs. \cite{Liu:2013xt,Alberte:2016izw}, such a
low-scale inflation is regarded as current accelerated expansion
with $H^2_{inf1}\sim \Lambda$.
The consistent joint of an NEC-preserving spacetime to an NEC-violating phase is also explored in Ref. \cite{Rubakov:2013kaa}, see also \cite{Elder:2013gya}.

In this paper, we investigate the possibility of a short NEC
violation during the NEC-preserving inflation. In this scenario, the low-scale inflation prior to NEC violation is responsible for the density
perturbation on large scales. We calculate the corresponding
primordial GW spectrum. Specially, we highlighted that for the
multi-stage (NEC-preserving) inflation, if the NEC violations happened intermittently, a Great Wall-like landscape of primordial
GW spectrum at the full frequency-band will present.

\section{Our scenario }\label{NECinf}

\subsection{Intermittent NEC violation during inflation} \label{NEC-model}

\begin{figure}[htbp]
    \includegraphics[scale=2,width=0.55\textwidth]{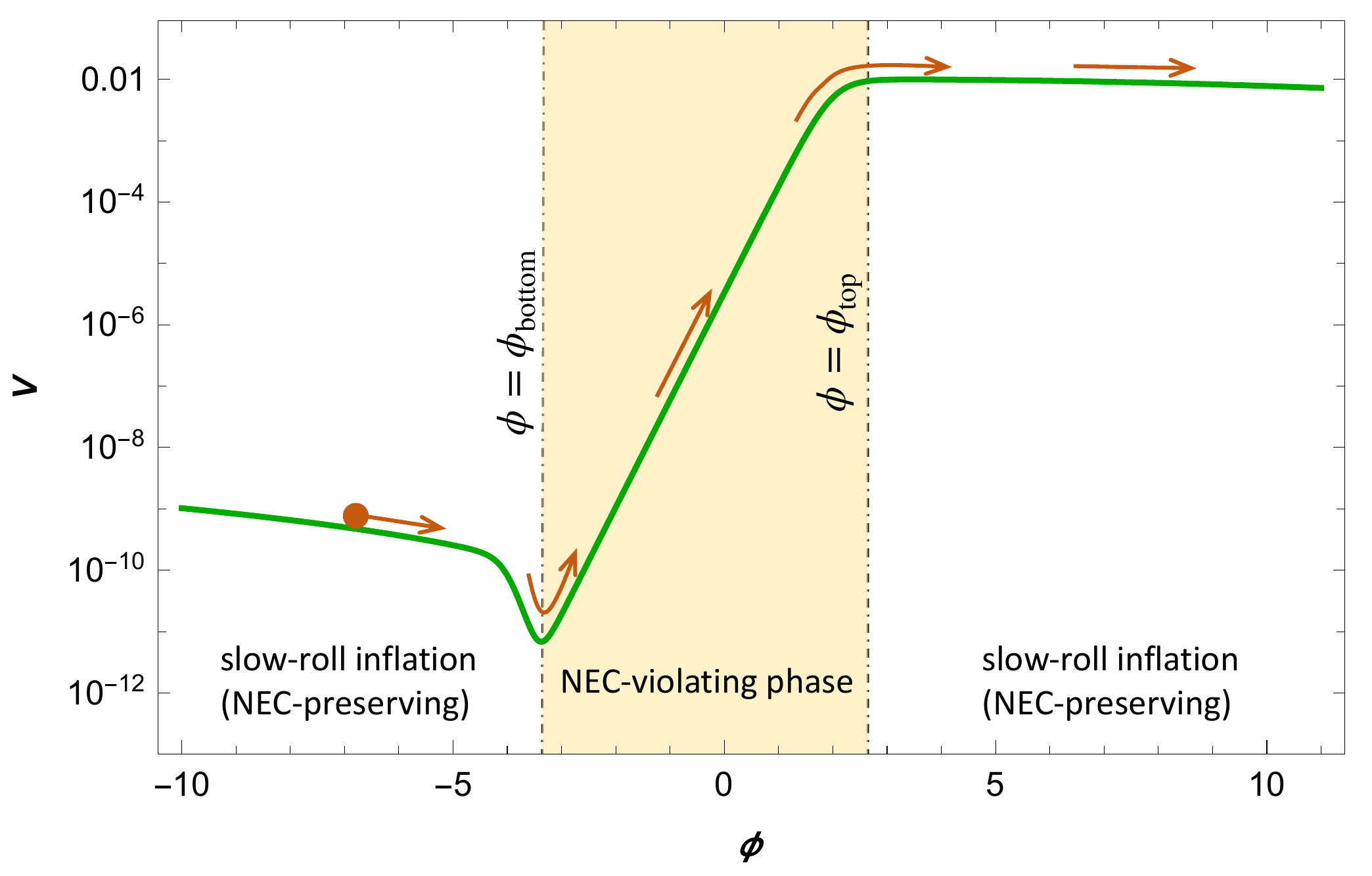}
\caption{A sketch of our scenario: a slow-roll (NEC-preserving)
inflation occurred before the NEC-violating phase, which is
subsequently followed by the slow-roll (NEC-preserving) inflation
with a higher scale $H_{inf2}\gg H_{inf1}$. The potential $V(\phi)$ given by Eq. (\ref{V}) is
plotted with logarithmic coordinates on the vertical axis.} \label{fig:figV}
\end{figure}

In our scenario (see Fig. \ref{fig:figV}), initially the field $\phi$ (canonical scalar field) slowly rolls down a nearly flat potential, i.e., ${\dot\phi}^2\ll V(\phi)\approx V_{inf1}$, which results in the slow-roll
(NEC-preserving) inflation.
In the NEC-violating phase, $\phi$ climbs up the potential rapidly so that
${\dot H}>0$. After $\phi$ arrives at another nearly flat region of the potential but with higher energy, i.e.,
$V_{inf2}\gg V_{inf1}$, the
slow-roll inflation restarts again.

We present a model as follows: \ba\label{genesis-action} &\,&
S=\int d^4x\sqrt{-g}\Big[{M_p^2\over 2}R - M_p^2
    g_1(\phi)X/2 + g_2(\phi)X^2/4 -M_p^4 V(\phi)  \Big] \,, \ea
where $X=\nabla_{\mu}\phi\nabla^\mu\phi$. Here, the Galileon
operator $\Box \phi=\nabla_{\mu}\nabla^{\mu}\phi$ is not required,\footnote{See e.g., Ref. \cite{Tahara:2020fmn}, for the case in which $X\Box \phi$ instead of $X^2$ is used to realize the NEC violation.} see also
e.g. \cite{Cai:2017dyi,Buchbinder:2007ad,Koehn:2015vvy}. The
corresponding background equations are \ba \label{eqH} &\,& 3 H^2
M_p^2=\frac{M_p^2}{2} g_1\dot{\phi }^2 +\frac{3}{4} g_2 \dot{\phi
}^4 +M_p^4 V \,,
\\&\,&
\dot{H} M_p^2= -\frac{M_p^2}{2} g_1 \dot{\phi }^2 -\frac{1}{2} g_2
\dot{\phi }^4\,,  \label{dotH}
\\&\,&
0= \lf(g_1 +\frac{3 g_2 \dot{\phi }^2 }{M_p^2}\rt)\ddot{\phi}
+3 g_1 H\dot{\phi } +\frac{1}{2}g_{1,\phi }
\dot{\phi }^2 
+\frac{3 g_2 H\dot{\phi }^3}{M_p^2}  +\frac{3 g_{2,\phi } \dot{\phi }^4}{4M_p^2}
+M_p^2 V_{,\phi }\,, \label{eomphi} \ea where
``$_{,\phi}=d/d\phi$." Only two of Eqs. (\ref{eqH})-(\ref{eomphi})
are independent.

In the NEC-preserving regimes, we require $g_1(\phi)=1$,
$g_2(\phi)=0$ and the potential is nearly flat (see Fig. \ref{fig:figV}), so that the scalar
filed $\phi$ is canonical and the slow-roll inflation ($0<\epsilon
=-{\dot H}/H^2\ll 1$) can happen. In Fig. \ref{fig:figV}, $V_{inf1}\simeq
m^2\phi^2/2$, $V_{inf2}\simeq
\lambda[1-\frac{(\phi-\phi_1)^{2}}{\sigma^{2}}]^2$ and $V_{inf2}\gg
V_{inf1}$.

In the NEC-violating regime, we set $g_1(\phi)\approx - {f_1
    e^{2\phi} \over 1+f_1 e^{2\phi} }<0$ and $g_2(\phi)=f_2$ with
$f_{1,2}$ being dimensionless constants. The coefficient of
$\ddot{\phi}$ in Eq. (\ref{eomphi}), i.e.,
$g_1+3g_2\dot{\phi}^2/M_p^2$, is positive throughout so that there is no
ghost instability, see Appendix \ref{Sec:app}. The scalar field $\phi$ will climb up the potential rapidly ($H\ll\dot{\phi}<M_p$) and arrive at the flat region $V=V_{inf2}$, as long as the condition
$\frac{1}{2}g_{1,\phi } \dot{\phi }^2+M_p^2 V_{,\phi }<0$ lasts for
sufficiently long time. We require $\epsilon \ll -1$, i.e., $H^2\ll
{\dot H}$. According to Eq. (\ref{eqH}), considering
${\dot\phi}^2\gg V,\,H^2$, we have $\frac{M_p^2}{2} g_1\dot{\phi
}^2 +\frac{3}{4} g_2 \dot{\phi }^4\approx 0$, which suggests
$e^{2\phi}\sim\dot{\phi}^2$ for $\phi<0$. Thus, $\dot \phi$ is
approximately \be \label{dotphi} \dot{\phi}\simeq {1\over
    (t_*-t)}\,,\quad t<t_*\,. \ee According to Eq. (\ref{dotH}), we
have $\dot {H}\sim \dot{\phi}^4$; hence, \be \label{HH} H\sim {1\over
    (t_*-t)^3}+{\rm const..} \ee When $t\ll t_*$, we have $H\simeq
{\rm const.}=H_{inf1}$, which suggests that the NEC-violating phase has
the chance to start after a slow-roll inflation.

As a phenomenological example, we set \ba g_1(\phi)&=&
{1\over 1+e^{q_2(\phi-\phi_3)}} -{f_1 e^{2\phi} \over 1+f_1
e^{2\phi} }+{2\over 1+e^{-q_1(\phi-\phi_0)}}\,,\label{g1}
\\
g_2(\phi)&=& {f_2\over 1+e^{-q_2(\phi-\phi_3)}} {1\over
    1+e^{q_3(\phi-\phi_0)}}\,,
\label{g3}
\\ V(\phi)&=&{1\over2}m^2\phi^2 {1\over
    1+e^{q_2(\phi-\phi_2)}}+
\lambda\left[1-\frac{(\phi-\phi_1)^{2}}{\sigma^{2}}\right]^{2}{1\over
    1+e^{-q_4(\phi-\phi_1)}}\,, \label{V}
\ea
where $\lambda$, $m$, $f_{1,2}$ and $q_{1,2,3,4}$ are positive
constants. We require that $\phi_3<\phi_2<0<\phi_1<\phi_0$. Here,
for $\phi\ll \phi_3$, we have $g_1= 1$, $g_2= 0$ and
$V=V_{inf1}\simeq m^2\phi^2/2$, while for $\phi\gg \phi_0$, we
have $g_1= 1$, $g_2= 0$ and $V=V_{inf2}\simeq \lambda
[1-\frac{(\phi-\phi_1)^{2}}{\sigma^{2}}]^{2}$.

We solve Eqs. (\ref{dotH}) and (\ref{eomphi}) numerically. The initial value of $H$ is set as $H_{\rm ini}\simeq H_{inf1}=1.29\times
10^{-5}M_p$ at $t=t_{\rm ini}=0$, so that the ``$inf1$'' is responsible for the scalar
perturbations on the CMB band, which indicates that $P_{s} \simeq \frac{1}{2 M_{p}^{2}
    \epsilon_{inf1}}\left(\frac{H_{inf1}}{2
    \pi}\right)^{2}\approx 2.1\times 10^{-9}$ for $\epsilon_{inf1}=0.001$.

\begin{figure}[htbp]
    \centering
\subfigure[ ]
{\includegraphics[width=0.46\textwidth]{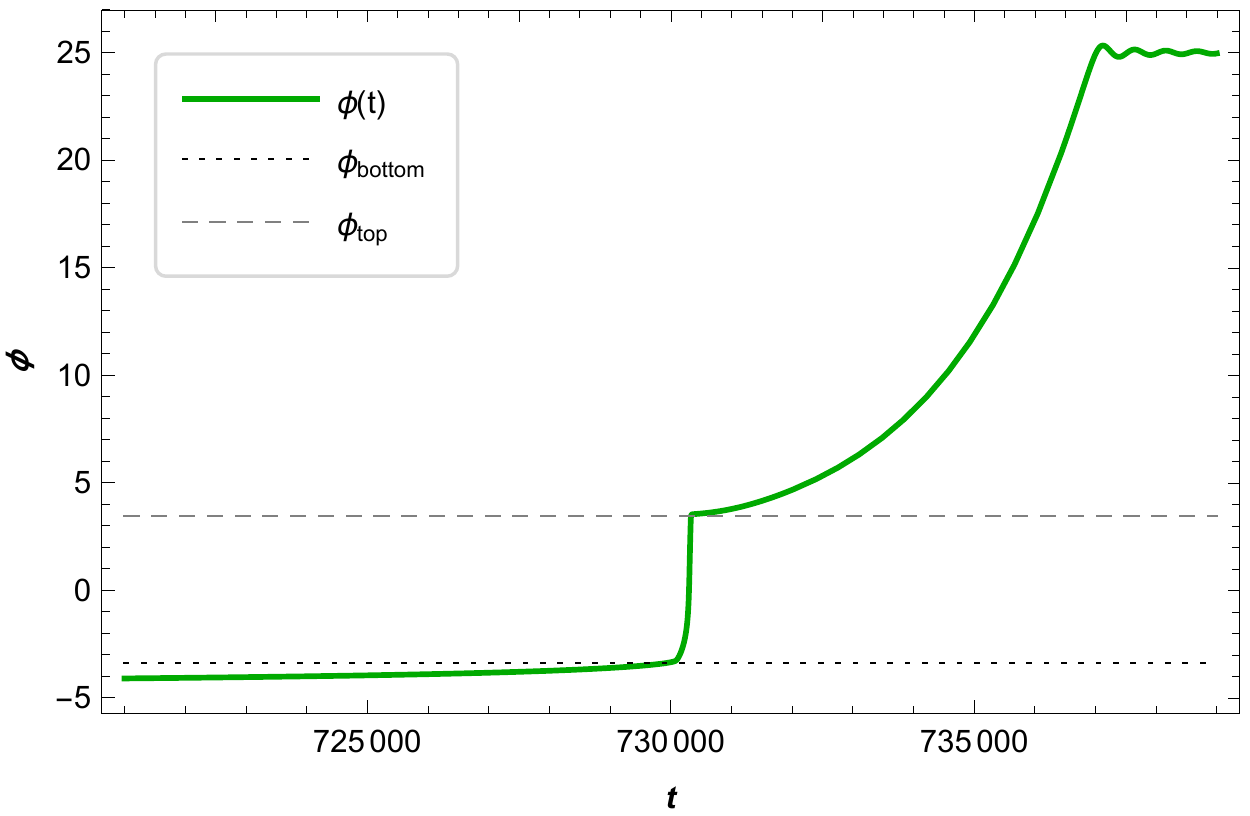}\label{fig:figphi-num}}
    \quad
\subfigure[ ]
{\includegraphics[width=0.47\textwidth]{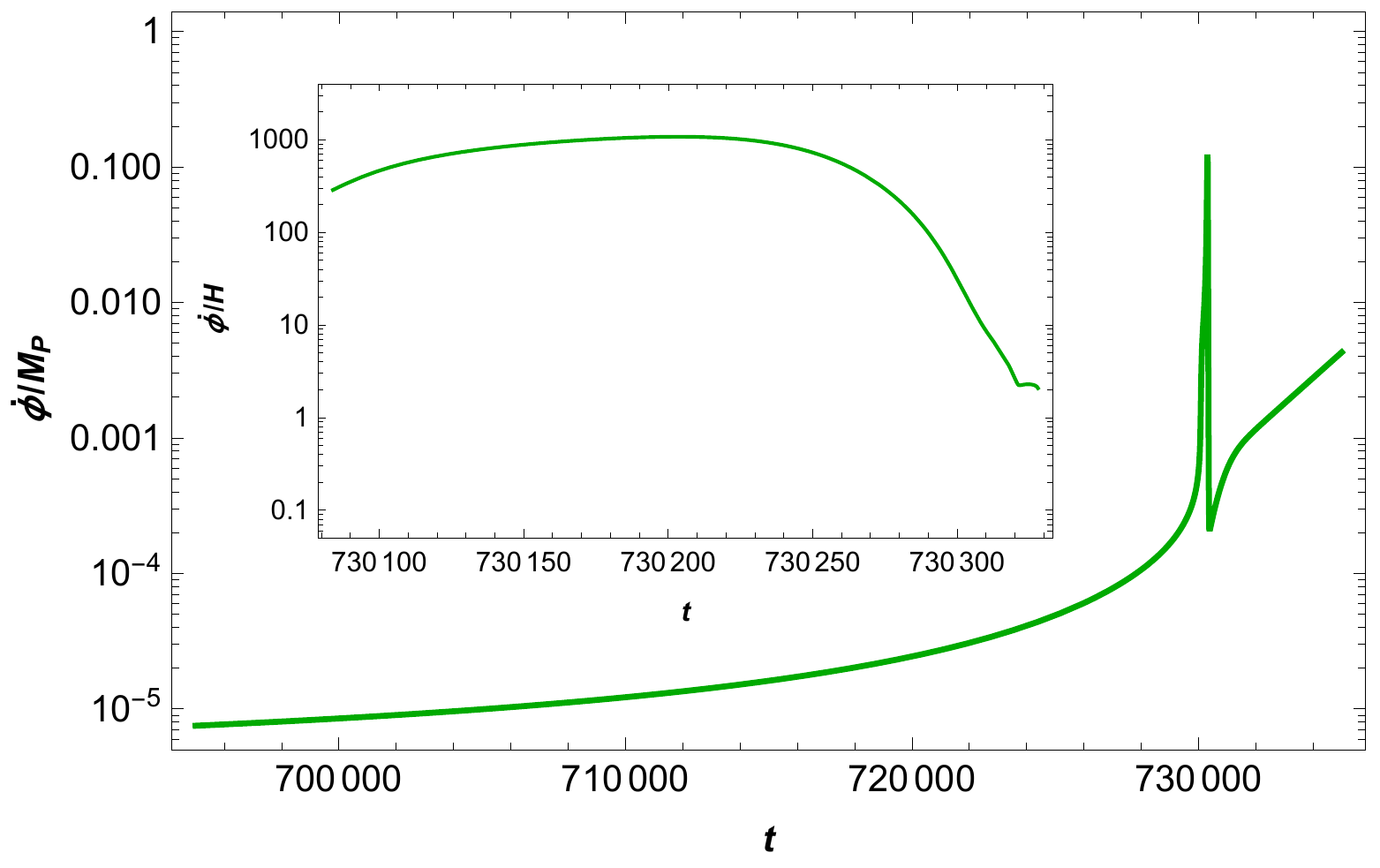}\label{fig:dotfigphi-num}}
\caption{Left: the evolution of $\phi$ with respect to $t$. Here,
$\phi_{\rm bottom}$ and $\phi_{\rm top}$ correspond to the bottom
and top of the potential in Fig. \ref{fig:figV}, respectively.
Right: the evolution of $\dot{\phi}/M_p$ and $\dot{\phi}/H$. During the NEC-violating
phase, which approximately corresponds to $730084<t<730328$,
$H\ll\dot{\phi}<M_p$ is satisfied. We set $\phi(t_{\rm ini})=-7$,
$\dot{\phi}(t_{\rm ini})=0$, $t_{\rm ini}=0$, $\phi_0=3.2$, $\phi_1=2$, $\phi_2=-4$, $\phi_3=-4.38$,
 $q_1=10$, $q_2=6$, $q_3=10$, $q_4=4$,
$f_1=1$, $f_2=40$, $\lambda=0.01$, $\sigma=23$ and $m=-4.5\times
10^{-6}$.} \label{phidotphi}
\end{figure}

\begin{figure}[htbp]
    \centering
\subfigure[ ]
{\includegraphics[width=0.48\textwidth]{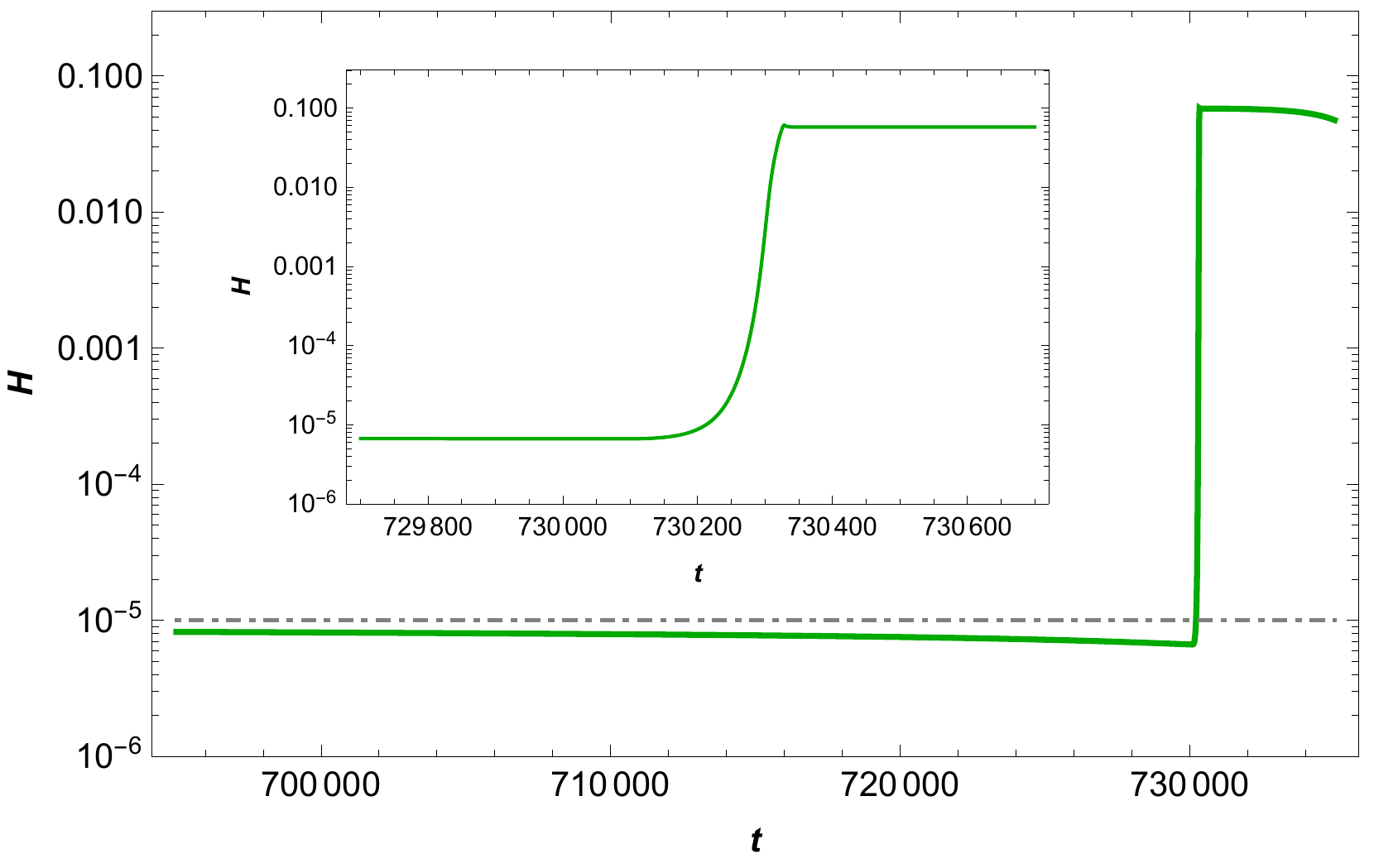}\label{fig:figH-num}}
    \quad
\subfigure[ ]
{\includegraphics[width=0.47\textwidth]{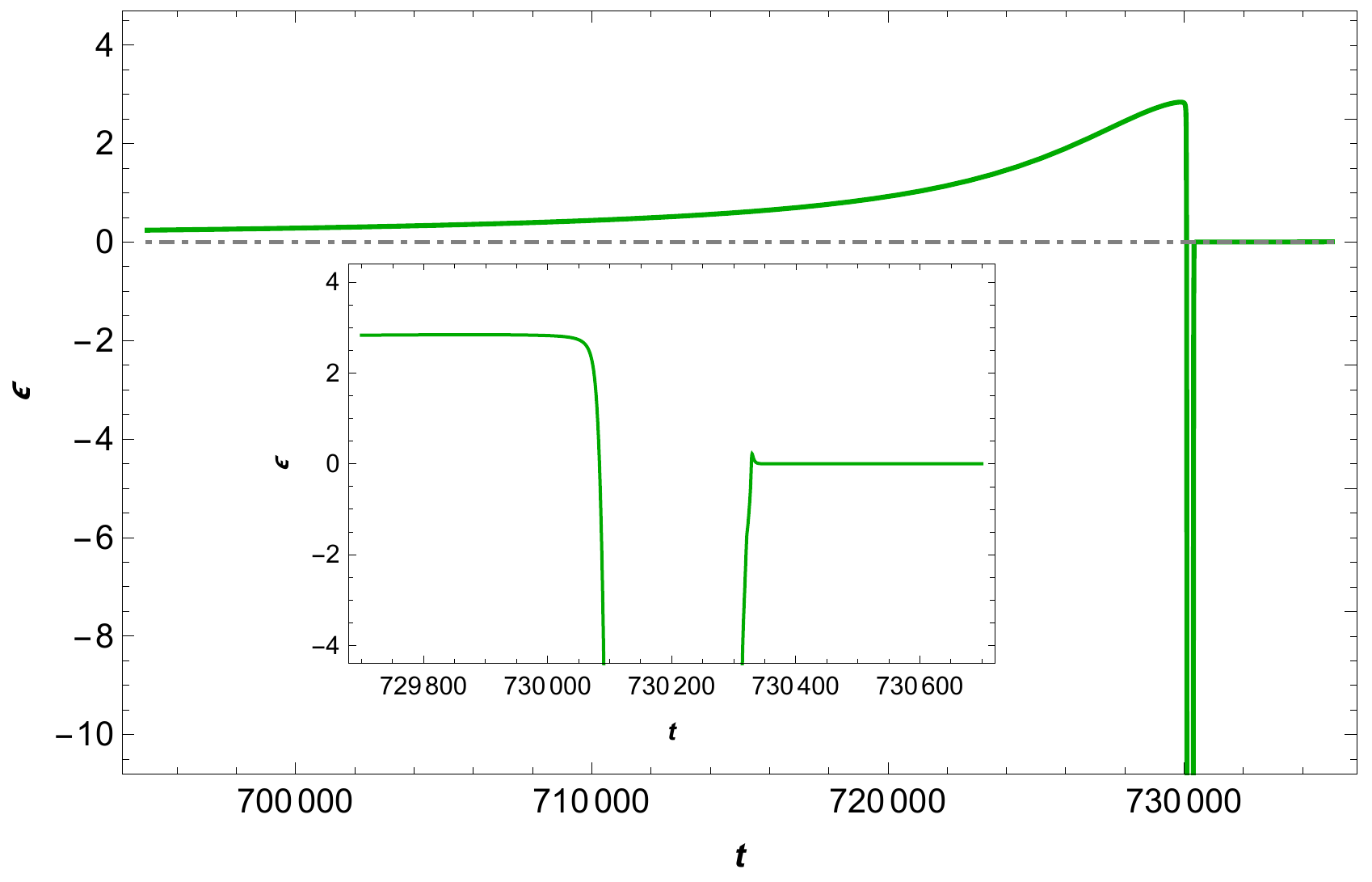}\label{fig:figepsilon-num}}
\caption{Left: the evolution of $H$. Right: the evolution of
$\epsilon=-\dot{H}/H^2$. We have $0<\epsilon\ll1$ during slow-roll inflations and $\epsilon\ll-1$ during NEC-violating super-inflation. }\label{Hepsilon}
\end{figure}

We plot the evolutions of $\phi$ and $\dot \phi$ in
Fig. \ref{phidotphi}. We can see that in the slow-roll (NEC-preserving)
regimes, ${\dot\phi}\ll H$, the field $\phi$ rolls slowly. In the
NEC-violating regime, $H\ll{\dot\phi}< M_p$, so that the field can
rapidly climb up the potential $V=V_{inf2}$. We plot the
evolutions of $H$ and $\epsilon$ in Fig. \ref{Hepsilon}. During the
slow-roll (NEC-preserving) phases ($0<\epsilon\ll1$), we have
$H\simeq {\rm const.}$, which is intervened by an NEC-violating phase
(${\dot H}>0$ and $\epsilon\ll-1$).
Due to the NEC-violating evolution, we have
\be {H_{inf2}/
    H_{inf1}}\simeq 10^3\gg 1.
\ee

\subsection{Primordial GW spectrum}

In this subsection, we calculate the spectrum of primordial
GWs. Generally, for the tensor perturbation $\gamma_{ij}$, we
have\footnote{Here, the propagating speed of GWs is $c_T=1$ (or
see
e.g., \cite{Cai:2015yza,Cai:2016ldn,Giovannini:2015kfa,Giovannini:2018zbf,Giare:2020vss,Giare:2020plo}).}
\be S^{(2)}_{\gamma}={M_p^2\over8}\int d^4xa^3 \lf[
\dot{\gamma}_{ij}^2 -{(\partial_k\gamma_{ij})^2\over
    a^2}\rt]\,.\label{tensor-action}
\ee In the momentum space, we have \be
\gamma_{ij}(\tau,\mathbf{x})=\int \frac{d^3k}{(2\pi)^{3}
}e^{-i\mathbf{k}\cdot \mathbf{x}} \sum_{\lambda=+,\times}
\hat{\gamma}_{\lambda}(\tau,\mathbf{k})
\epsilon^{(\lambda)}_{ij}(\mathbf{k}), \ee where
$\hat{\gamma}_{\lambda}(\tau,\mathbf{k})=
\gamma_{\lambda}(\tau,k)a_{\lambda}(\mathbf{k})
+\gamma_{\lambda}^*(\tau,-k)a_{\lambda}^{\dag}(-\mathbf{k})$,
$\epsilon_{ij}^{(\lambda)}(\mathbf{k})$ satisfies
$k_{j}\epsilon_{ij}^{(\lambda)}(\mathbf{k})=0$,
$\epsilon_{ii}^{(\lambda)}(\mathbf{k})=0$,
$\epsilon_{ij}^{(\lambda)}(\mathbf{k})
\epsilon_{ij}^{*(\lambda^{\prime}) }(\mathbf{k})=\delta_{\lambda
    \lambda^{\prime} }$ and $\epsilon_{ij}^{*(\lambda)
}(\mathbf{k})=\epsilon_{ij}^{(\lambda) }(-\mathbf{k})$;
$a_{\lambda}(\mathbf{k})$ and
$a^{\dag}_{\lambda}(\mathbf{k}^{\prime})$ satisfy $[
a_{\lambda}(\mathbf{k}),a_{\lambda^{\prime}}^{\dag}(\mathbf{k}^{\prime})
]=\delta_{\lambda\lambda^{\prime}}\delta^{(3)}(\mathbf{k}-\mathbf{k}^{\prime})$.
The equation of motion for $\gamma_{\lambda}(\tau,k)$ is \be \frac{d^{2}
u_k}{d \tau^{2}}+\left(k^{2}-\frac{a^{\prime \prime}}{a}\right)
u_k=0\,,\label{eq:eomu} \ee where
$u_k=\gamma_{\lambda}(\tau,k)aM_p/2$ and $\tau=\int a^{-1}dt$.

In Sec. \ref{NEC-model}, we have shown that a model of our scenario can be constructed.
Here, for simplicity, we assume that $\epsilon\approx {\rm const.}\ll -1$ during the NEC-violating phase, which requires some delicate design of $g_1$, $g_2$ and $V$. Around the beginning or the end of the NEC-violating phase, the detailed variation of $\epsilon$ may make some model-dependent contributions. However, for our purpose, the simplification will not make a qualitative difference.

Henceforth, we assume that
the epoch of ``inflation"  consists of different phases with
$\epsilon_j=-{\dot H}_j/H^2_j={3\over 2}(1+w_j)\simeq {\rm const.}$,
where $w_j$ is the state parameter. We have \cite{Cai:2015nya} \be
a_{j}(\tau)\sim
\left({\tau}_{R,j}-\tau\right)^{\frac{1}{\epsilon_{j}-1}}, \ee for
the $j$-th phase, where ${\tau}_{R,j} =\tau_{j}-
(\epsilon_{j}-1)^{-1} \mathcal{H}^{-1}(\tau_j)$ and $a(\tau_j)$ is
set by requiring the continuity of $a$ at the end of phase $j$
(i.e., $\tau=\tau_j$). As a result, we have \be \frac{a_j^{\prime
        \prime}}{a_j}={\nu_j^2-{1/4} \over
    {(\tau-\tau_{R,j})^2}}\,,\label{zppbz02} \ee where
$\nu_j={3\over2}\lf|{1-w_j\over 1+3 w_j}\rt|$. Regarding the
phases $j$ and $j+1$ as adjacent phases, we have the solutions to
Eq. (\ref{eq:eomu}) as \ba u_{k,j}(\tau)&=&{\sqrt{\pi
        (\tau_{R,j}-\tau)}\over
    2}\lf\{\alpha_{j}H_{\nu_{j}}^{(1)}[k(\tau_{R,j}-\tau)]+\beta_{j}H_{\nu_j}^{(2)}[k(\tau_{R,j}-\tau)]
\rt\} ,\,\, (\tau<\tau_{j})\,,
\\
u_{k,{j+1}}(\tau)&=&{\sqrt{\pi (\tau_{R,{j+1}}-\tau)}\over
    2}\Big\{\alpha_{j+1}H_{\nu_{j+1}}^{(1)}[k(\tau_{R,j+1}-\tau)]
\nn\\
&\,&\qquad\qquad\qquad\quad
+\beta_{j+1}H_{\nu_{j+1}}^{(2)}[k(\tau_{R,j+1}-\tau)] \Big\}
,\qquad\qquad\qquad (\tau>\tau_{j})\,, \label{solution}\ea
respectively, where $\alpha_{j(j+1)}$ and $\beta_{j(j+1)}$ are
$k$-dependent coefficients. Using the matching conditions
$u_{k,j}(\tau_{j+1})=u_{k,j+1}(\tau_{j+1})$ and
$u_{k,j}'(\tau_{j+1})=u_{k,j+1}'(\tau_{j+1})$, we have \ba \left(
\begin{array}{ccc} \alpha_{j+1}\\ \beta_{j+1}
\end{array}\right)
&=& {\cal M}^{(j)}
\left(\begin{array}{ccc} \alpha_{j}\\
    \beta_{j}\end{array}\right)\,,\quad {\rm where} \quad
{\cal M}^{(j)} = \left(\begin{array}{ccc}
    {\cal M}^{(j)}_{11}&{\cal
        M}^{(j)}_{12}\\
    {\cal M}^{(j)}_{21}&{\cal M}^{(j)}_{22}\end{array}\right)\,,
 \label{Mmetric}
\ea see Refs. \cite{Cai:2015nya,Cai:2019hge} for the matrix
elements of ${\cal M}^{(j)}$. The information of the $1,2\cdots
j$-th phases of the Universe has been encoded fully in the Bogoliubov
coefficients $\alpha_{j+1}$ and $\beta_{j+1}$. We set the initial
state as the Bunch-Davies vacuum (see also
\cite{Piao:2003zm,Piao:2005ag,Liu:2013kea,Qiu:2015nha,Cai:2017pga}
for pre-inflationary bounce), i.e., $u_k= \frac{1}{\sqrt{2 k}
}e^{-i k\tau}$. Thus, $|\alpha_1|=1$, $|\beta_1|=0$.

In the following, we focus on the scenario in Fig. \ref{fig:figV}.
Regarding ``$inf1$'', NEC-violating and ``$inf2$'' phases as the
$j=1,2,3$-th phases, respectively, we have \be
u_{k,3}(\tau)={\sqrt{\pi (\tau_{R,3}-\tau)}\over
2}\lf\{\alpha_{3}H_{3/2}^{(1)}[k(\tau_{R,3}-\tau)]+\beta_{3}H_{3/2}^{(2)}[k(\tau_{R,3}-\tau)]
\rt\} ,\ee where $\nu_3\simeq 3/2$ for de Sitter
expansion. On super-horizon scale, we have $
H^{(1)}_{3/2}(-k\tau)=-H_{3/2}^{(2)}(-k\tau)\overset{-k{\tau}\rightarrow0}
\approx-i \sqrt{2/(-\pi k^3\tau^3)}$. The resulting spectrum of
primordial GWs is \be P_T=\frac{4k^3}{\pi^2
M_p^2}\cdot\frac{\lf|u_{k,3}
    \rt|^2}{ a^2}= P_{T,inf2} \lf|\alpha_3 - \beta_3
\rt|^2\,,\label{eq:PT} \ee where \ba \left(
\begin{array}{ccc} \alpha_{3}\\ \beta_{3}
\end{array}\right)
&=& {\cal M}^{(2)} {\cal M}^{(1)}
\left(\begin{array}{ccc} \alpha_{1}\\
    \beta_{1}\end{array}\right)\,,
\label{Mmetric-2} \ea  $P_{T,inf2}={2H_{inf2}^2 \over M_p^2
\pi^2}$, ${\cal M}^{(1)}$ and ${\cal M}^{(2)}$ are given by Eq. (\ref{Mmetric}). The information of ``$inf1$'', NEC-violating and ``$inf2$''
phases has been encoded in the Bogoliubov coefficients $\alpha_3$
and $\beta_3$.

\subsection{Primordial GWB at low-frequency band}

It is interesting to connect $P_T$ in (\ref{eq:PT}) with the
observations of stochastic GWB at low-frequency bands. The
BICEP/Keck+Planck bound at CMB band is $r\lesssim 0.06$
\cite{Ade:2018gkx}, which corresponds to $P_T\lesssim 10^{-10}$.
The analysis result of NANOGrav 12.5-yr data
\cite{Arzoumanian:2020vkk}, if regarded as the stochastic GWB (see
inspired studies
e.g. \cite{Ellis:2020ena,Blasi:2020mfx,DeLuca:2020agl,Buchmuller:2020lbh,Addazi:2020zcj,Kohri:2020qqd,Ratzinger:2020koh,Samanta:2020cdk,Bian:2020bps,Vagnozzi:2020gtf,Neronov:2020qrl,Li:2020cjj,Sugiyama:2020roc,Liu:2020mru,Domenech:2020ers,Bhattacharya:2020lhc,Wong:2020yig,Kuroyanagi:2020sfw}),
suggests $\Omega_{GW}\sim 10^{-9}$ with the tilt $-1.5\lesssim
n_T\lesssim 0.5$, where
\be
\Omega_{GW}(\tau_{0})=\frac{k^{2}}{12
a_0^2H^2_0}P_{T}(k)\lf[\frac{3
\Omega_{{m}}j_1(k\tau_0)}{k\tau_{0}}\sqrt{1.0+1.36\frac{k}{k_{\text{eq}}}
+2.50\left( \frac{k}{k_{\text{eq}}}\right) ^{2}}\rt]^2,
\label{GW0}
\ee
is the energy density spectrum of GWs, see e.g. \cite{Turner:1993vb} (see also
\cite{Boyle:2005se,Zhao:2006mm,Kuroyanagi:2014nba,Liu:2015psa}).
Here, $1/k_{eq}$ is the comoving Hubble scale at matter-radiation equality,
$\Omega_m=\rho_m/\rho_c$ and $\rho_{{c}}=3H^{2}_0/\big(8\pi
G\big)$ is the critical energy density.

\begin{figure}[htbp]
    \centering
    \subfigure[ ] {\includegraphics[width=0.49\textwidth]{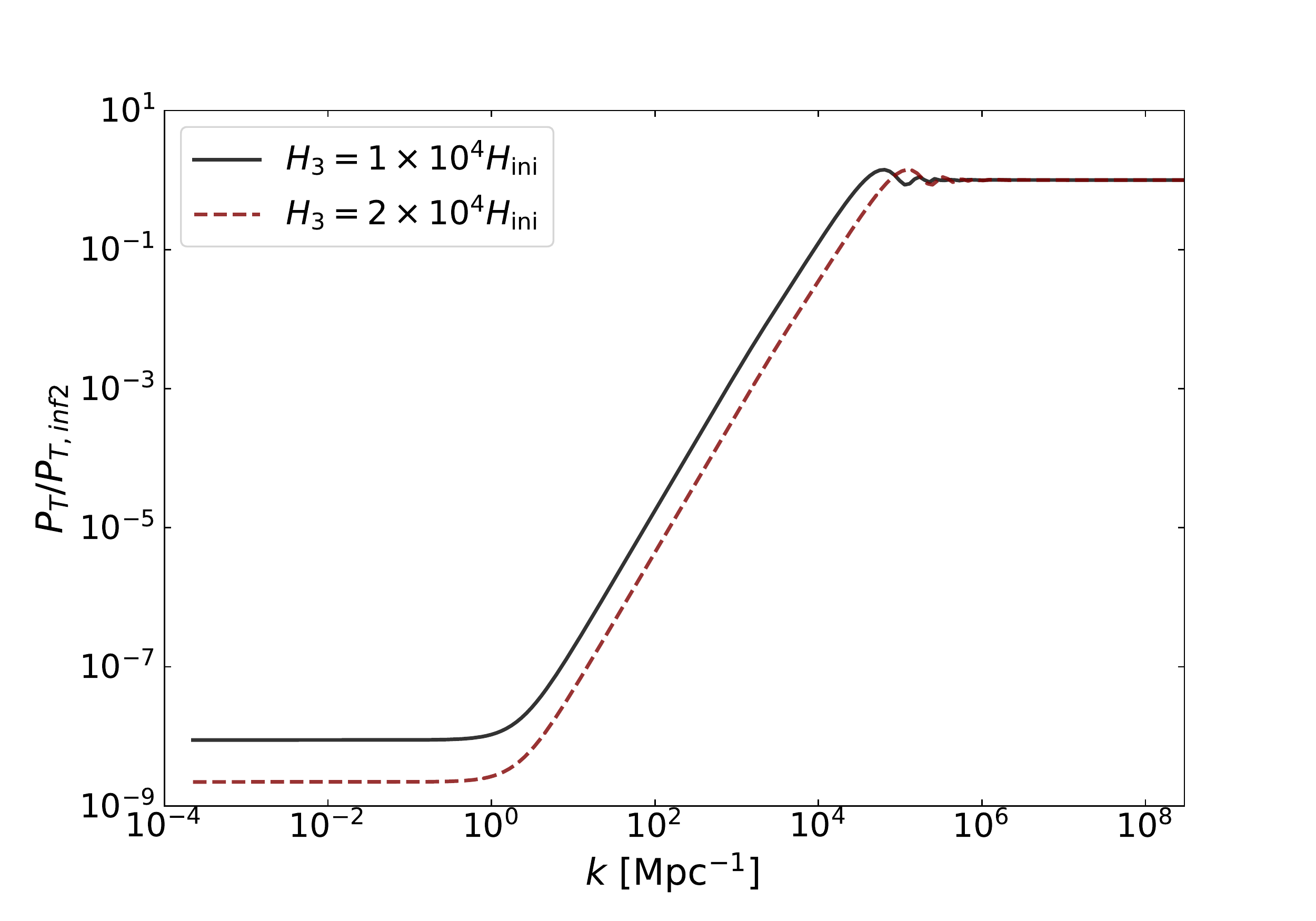}}
    \subfigure[  ] {\includegraphics[width=0.49\textwidth]{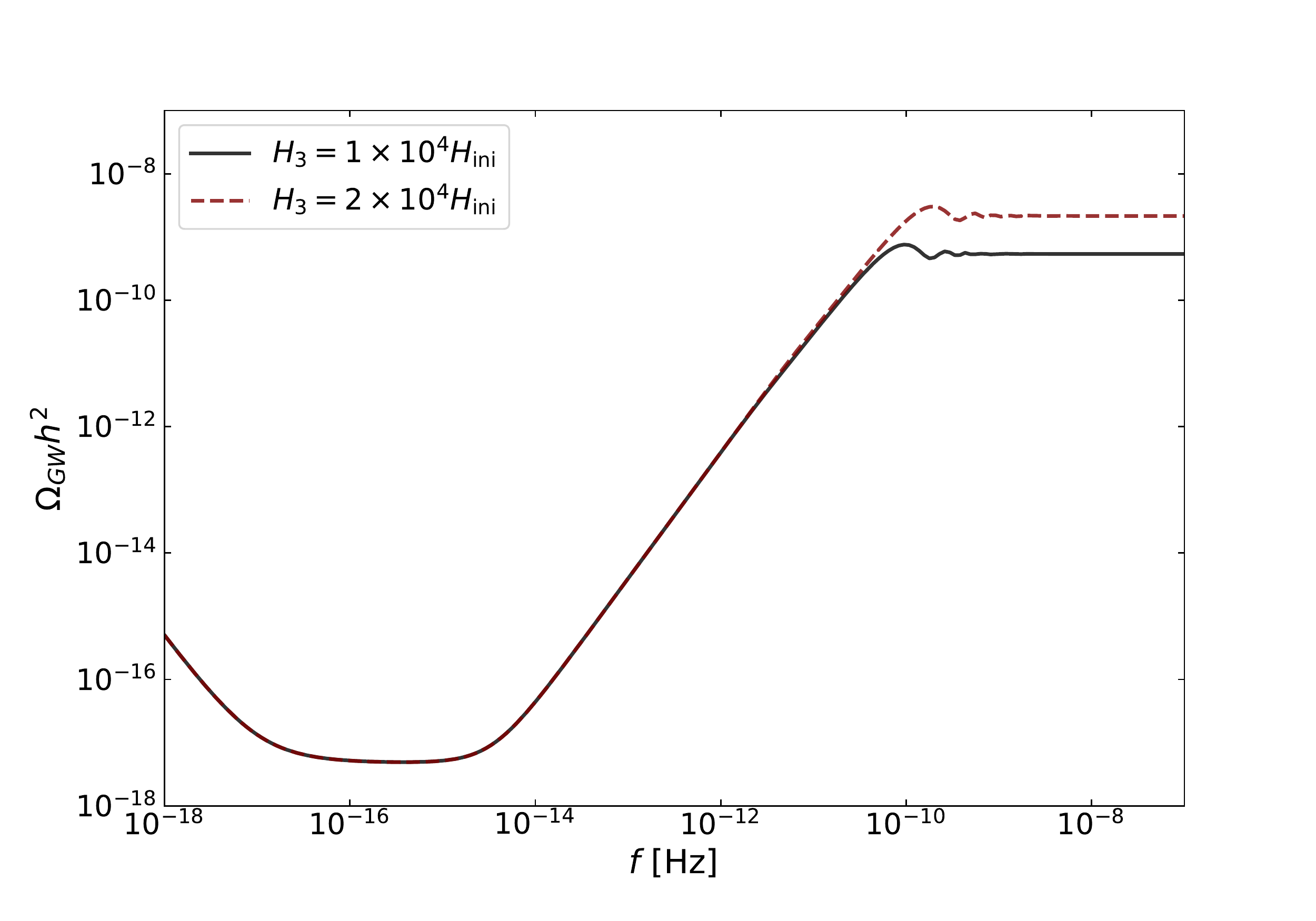}}
\caption{Left: $P_T$. Right: $\Omega_{\rm GW}h^2$, where
$h=H_0/({\rm km/s/Mpc})$. }\label{figOmegaGW}
\end{figure}

According to (\ref{eq:PT}), we plot $P_T$ and $\Omega_{GW}$ in
Fig. \ref{figOmegaGW} ($f=k/(2\pi a_0)$). We set $H_{1}=1.29\times
10^{-5}$, which corresponds to $P_{s} \sim 2.1\times 10^{-9}$ for
$\epsilon_{1}=0.001$ and $H_3\sim 10^{-2}$; $w_1\gtrsim -1$ and
$w_3\gtrsim -1$ for the slow-roll inflations, while $w_2\lesssim
-10$ for the NEC-violating phase. We see that the yielded power spectrum of primordial GWs has a nearly flat amplitude at the CMB band and also a
higher nearly flat amplitude at the PTA band. Here, the NEC-violating
regime contributes the upward section of $P_T$, in which the
spectrum has a blue tilt $n_T\simeq 2$ (since $\epsilon\ll -1$).
Therefore, our scenario not only explains the result reported by the
NANOGrav Collaboration, but also has a detectable signal $r\sim
0.01$ in the CMB.

\subsection{Multi-stage inflation with NEC violations}

The multi-stage inflation model (see e.g., earlier
Refs. \cite{Adams:1997de,Burgess:2005sb,Ashoorioon:2006wc,Ashoorioon:2008qr,Liu:2009pk}; see also e.g., \cite{DAmico:2020euu} for recent study), in which a
sequence of short inflations are interrupted by short periods of
decelerated expansions with $w>-1/3$, is interesting, since it
helps to make the EFT of inflation UV-complete
\cite{Obied:2018sgi,Bedroya:2019snp,Li:2019ipk,Berera:2019zdd,Dhuria:2019oyf,Torabian:2019zms}.
Usually, in such a scenario, a high-scale inflation is followed by
a sequence of low-scale inflations. However, it might be also
possible that a sequence of short inflations ($w\gtrsim -1$) are
interrupted by short periods of not only decelerated expansions
with $w>-1/3$ but also super-inflation or Genesis with $w<-1$, so
that the scales of subsequent short (NEC-preserving) inflations
might be higher, see e.g. Fig. \ref{fig:figV}.

According to Eqs. (\ref{solution}), (\ref{Mmetric}) and
(\ref{eq:PT}), for a multi-stage scenario of inflation in which a
sequence of short slow-roll inflations ($w\gtrsim -1$) are
interrupted by lots of short periods of expansions with $w>-1/3$
and $w<-1$ (NEC violation), we can write the spectrum $P_T$ of
primordial GWs as  
\be P_T=P_{T,l}^{\rm inf} \lf|\alpha_{l} -
\beta_{l} \rt|^2 ={2H_{l}^2 \over M_p^2 \pi^2}\lf|\alpha_{l} -
\beta_{l} \rt|^2\,, \label{multiPT} \ee where \ba \left(
\begin{array}{ccc} \alpha_{l}\\ \beta_{l}
\end{array}\right)
&=& \prod_{j=1}^{l} {\cal M}^{(j)}
\left(\begin{array}{ccc} \alpha_{1}\\
    \beta_{1}\end{array}\right)\,,
\label{Mmetric-3} \ea and `$l$' labels the last short slow-roll
inflations. The frequency band of stochastic GWB yielded is \be
10^{-18} \,{\rm Hz}\lesssim f\lesssim \exp{\lf(\sum_{j=1}^l
N_j\rt)}10^{-18}\,{\rm Hz}, \ee where $N_j\equiv \ln{a_{j,e}H_{j,e}\over
a_{j,ini}H_{j,ini}}$ is the $e$-folds number of the perturbation
modes passing through the $j$-th phase. Note that $N_j<0$ for the
decelerated expansion ($w>-1/3$).

\begin{figure}[htbp]
\centering
    \subfigure[ ]{\includegraphics[width=0.95\textwidth]{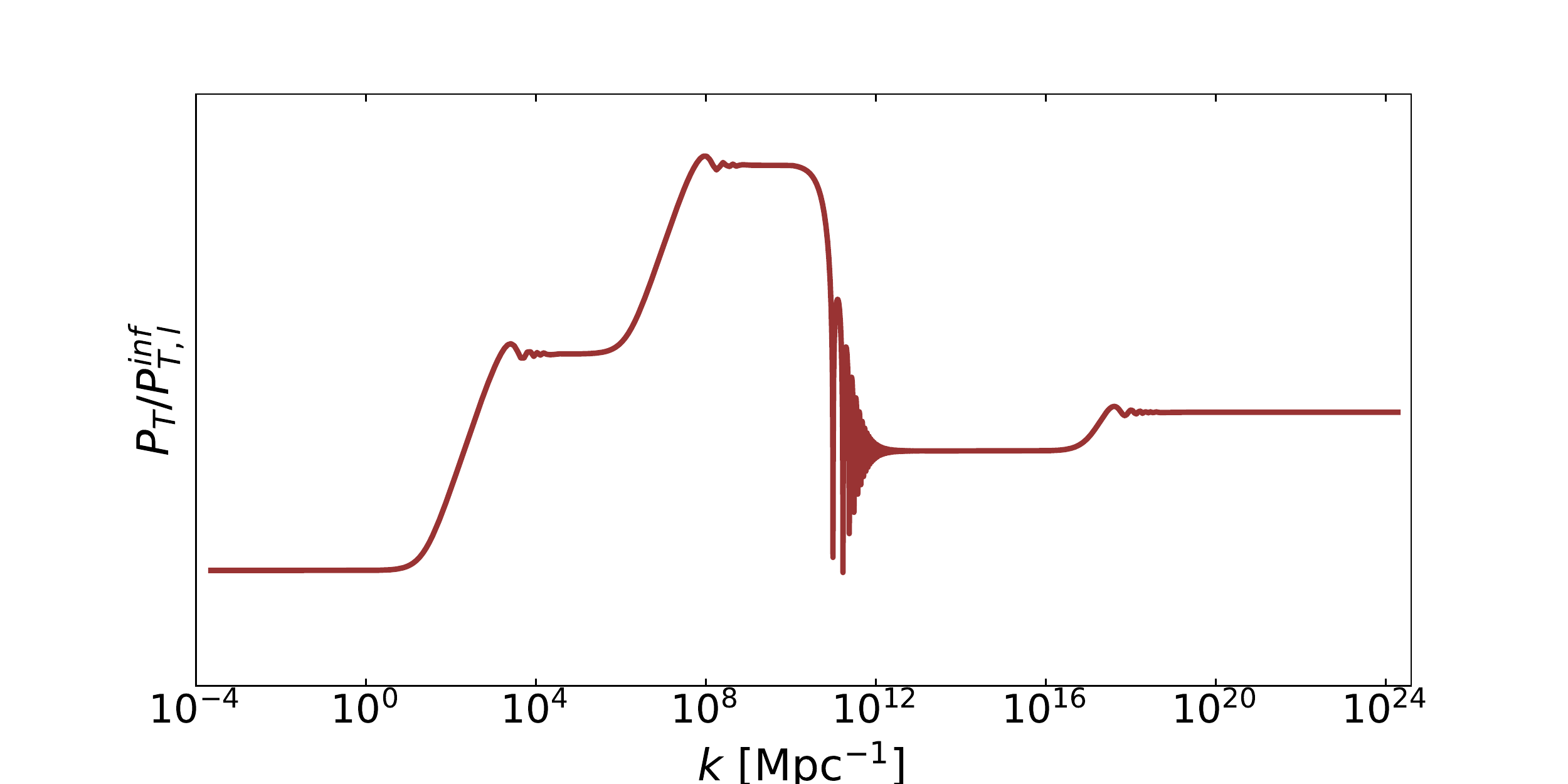}}
    \subfigure[ ]{\includegraphics[width=0.75\textwidth]{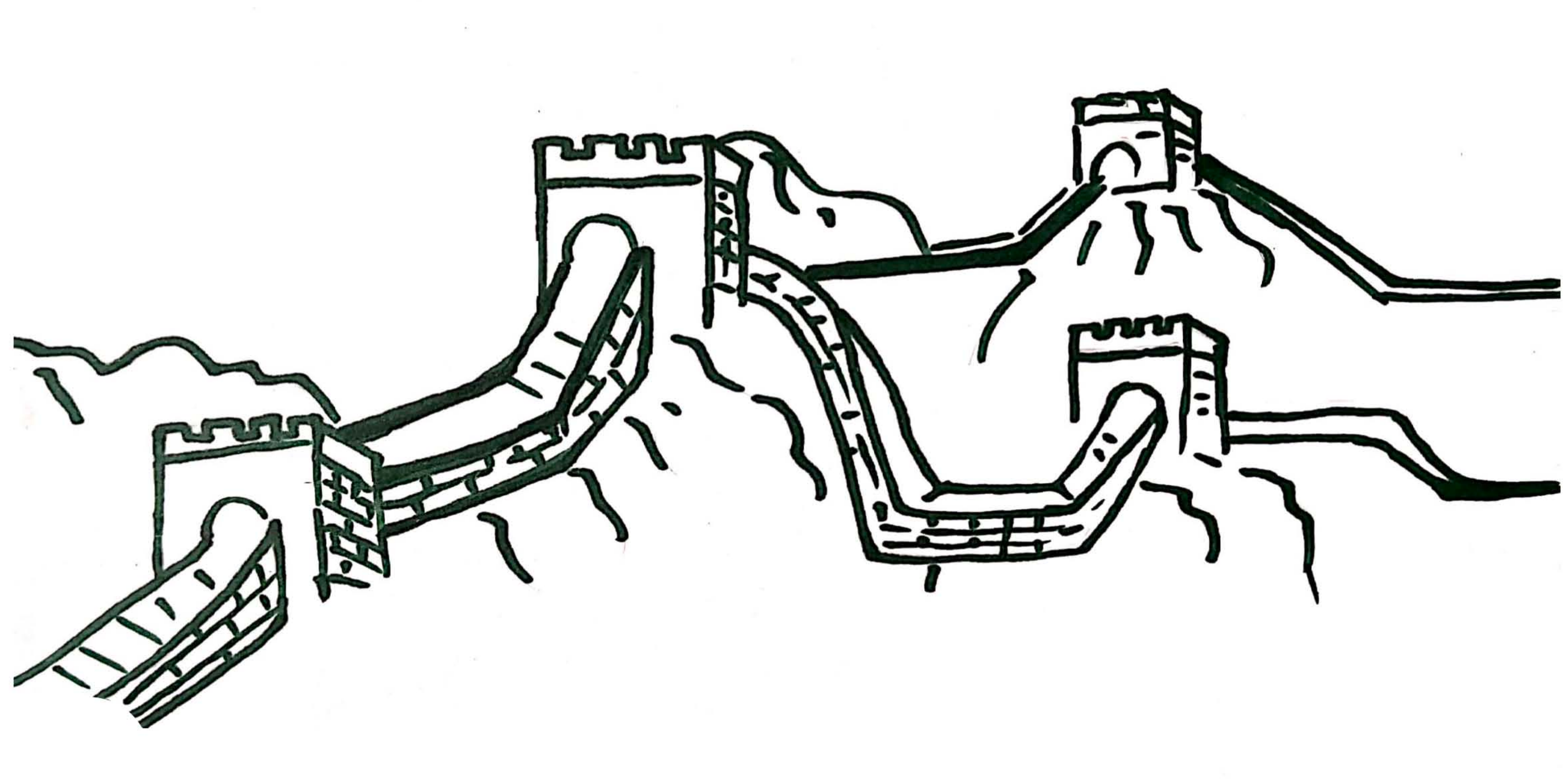}}
\caption{The spectrum $P_T/P_{T,l}^{inf}$ of primordial GWB
    yielded in a multi-stage scenario of inflation, in which a
    sequence of short slow-roll inflations ($j=1,3,5,7,9$) are
    interrupted by short periods of decelerated expansion ($j=6$) and
    NEC-violating expansion ($j=2,4,8$). We set the equation of state
    parameters $w_{1,3,5,7,9}\simeq -1$, $w_2=-15$, $w_{4,8}=-10$ and
    $w_6=1/3$. The frequency band of GW spans about $28$ orders. The
    lower panel is the Great Wall (sketched by Yu-Ze Piao). The
    panorama of $P_T$ looks like the Great Wall. When we climb up the
    Great Wall, we would see the beacon towers of different physics.}\label{fig:greatwall}
\end{figure}

According to (\ref{multiPT}), we
plot $P_T/P_{T,l}^{inf}$ in Fig. \ref{fig:greatwall}(a) for a
multi-stage scenario of inflation with short periods of slow-roll
inflations ($j=1,3,5,7,9$) with different $H_j$. The panorama of
$P_T$ at corresponding GW frequency band looks like the Great
Wall, see Fig. \ref{fig:greatwall}(b), in which the nearly flat
roads correspond to GWB yielded by short slow-roll
(NEC-preserving) inflations, the upward and downward slopes
correspond to the NEC-violating expansions ($w< -1$) and
decelerated expansions ($w>-1/3$), respectively. It is well-known
that each section of the Great Wall records a unique history.

\section{Conclusion}

The NEC violation in the primordial Universe will bring a blue-tilted
GWB. However, its implications to the GWB might be far richer than
expected. We presented a scenario, in which after a slow-roll
(NEC-preserving) inflation with $H\simeq H_{inf1}$ (responsible
for the density perturbation on large scales), the Universe goes through an NEC-violating period, which is followed again by
the slow-roll inflation but with $H_{inf2}\gg H_{inf1}$. We
calculated the power spectrum of the yielded primordial GWs. As
expected, the spectrum has an observable amplitude $P_T\sim
H_{inf1}^2$ ($n_T\simeq 0$) at the CMB band and a higher amplitude
$P_T\sim H_{inf2}^2$ ($n_T\simeq 0$) at the PTA band (compatible with recent NANOGrav result). Here, the NEC violation
responsible for the upward tilt of $P_T$ played an indispensable
role.

It is well-known that the detection of stochastic GWB will not only solidify our confidence in inflation but also offer us a probe to the physics of the early Universe. Though the model we
consider is simplified, it highlights an unexpected point that the
GWB yielded in the primordial Universe might have a unique landscape.
We explore the observable imprints of short NEC violations on
primordial GWB. It is especially highlighted that for the
multi-stage inflation, consisting of a sequence of short slow-roll
inflations ($w\gtrsim -1$) interrupted by lots of short period of
expansions with $w>-1/3$ and $w<-1$, we will have a Great Wall
spectrum of stochastic GWB, which might be detectable.

\textbf{Acknowledgments}

We thank Gen Ye for helpful discussion. Y. C. is funded by the
China Postdoctoral Science Foundation (Grant No. 2019M650810) and
the NSFC (Grant No. 11905224). Y. S. P. is supported by NSFC
Grants No. 12075246 and No. 11690021.

\appendix

\section{On stability of scalar perturbations}\label{Sec:app}

\begin{figure}[tb]
    \centering
    \subfigure[ ] {\includegraphics[width=0.48\textwidth]{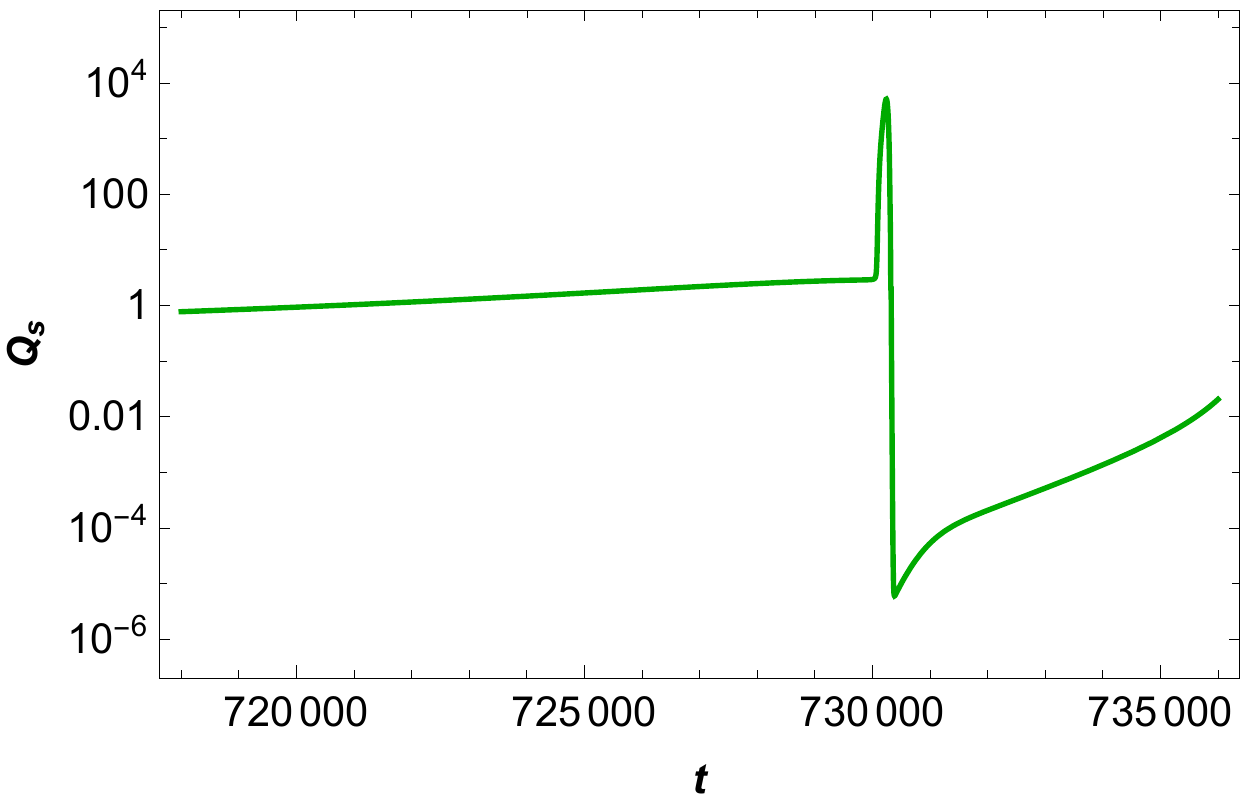}\label{fig:figQs-num}}
    \quad
    \subfigure[ ] {\includegraphics[width=0.47\textwidth]{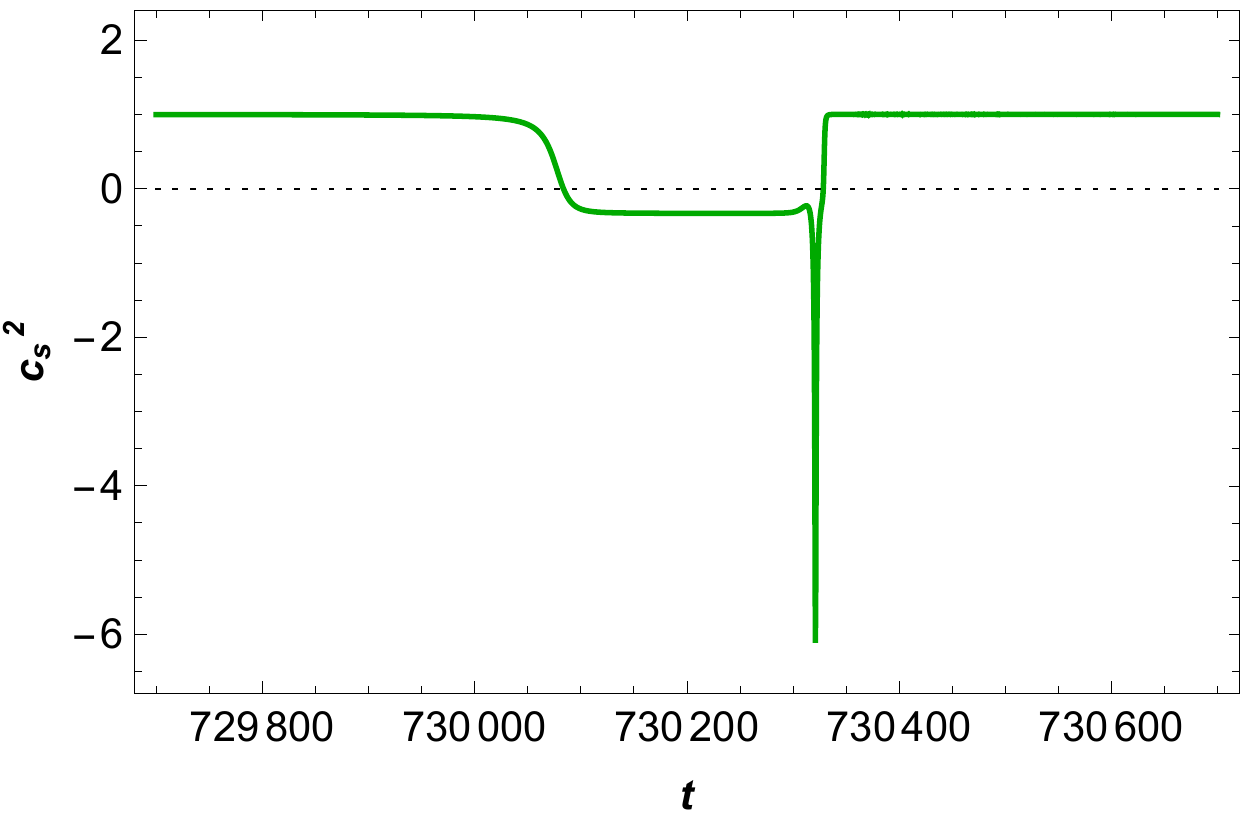}\label{fig:figcs2-num}}
\caption{Left: The evolution of $Q_s$, which is positive
throughout in our model. Right: The evolution of $c_s^2$, which is
negative during the NEC-violating phase (approximately
corresponding to $730084<t<730328$).}\label{Qscs}
\end{figure}

In the unitary gauge, for (\ref{genesis-action}), we have \be
\label{eft_action02} S^{(2)}_\zeta=\int d^4xa^3 Q_s\lf[
\dot{\zeta}^2-c_s^2{(\partial \zeta)^2\over a^2} \rt]\,, \ee where
\ba
    &\,&\label{eq:Qs} Q_s= \epsilon M_p^2+{ g_2\dot{\phi}^4 \over H^2}=
    {M_p^2\dot{\phi}^2 \over 2H^2}\lf(g_1+3 g_2{\dot{\phi}^2\over M_p^2}\rt)\,,\\
    &\,&\label{cs2}
    c_s^2={\epsilon M_p^2 \over Q_s}.
\ea  
Around the NEC violation, though $c_s^2<0$ (see Fig. \ref{Qscs}),
$c_s^2=1$ can be set with the higher-order derivative
(beyond-Horndeski) operators, see e.g.,
Refs. \cite{Cai:2016thi,Cai:2017tku,Cai:2017dyi,Cai:2017dxl,Ye:2019frg}
for related details. Here, $Q_s>0$ throughout. Therefore, the ghost and gradient instabilities can be cured in the EFT regime. It is interesting to ask whether there is a danger of instabilities from the higher-derivative terms in the perspective of the UV
theory. However, it is still unknown how to embed the EFT with such terms in a UV complete theory so far.  The related issues require further investigation in the future.

\bibliographystyle{utphys}

\bibliography{NEC-inf-Ref}

\providecommand{\href}[2]{#2}\begingroup\raggedright\begin{thebibliography}{10}

\bibitem{Starobinsky:1979ty}
A.~A. Starobinsky, ``{Spectrum of relict gravitational radiation and the early
  state of the universe},'' {\em JETP Lett.} {\bfseries 30} (1979) 682--685.

\bibitem{Rubakov:1982df}
V.~Rubakov, M.~Sazhin, and A.~Veryaskin, ``{Graviton Creation in the
  Inflationary Universe and the Grand Unification Scale},''
  \href{http://dx.doi.org/10.1016/0370-2693(82)90641-4}{{\em Phys. Lett. B}
  {\bfseries 115} (1982) 189--192}.

\bibitem{Rubakov:2014jja}
V.~Rubakov, ``{The Null Energy Condition and its violation},''
  \href{http://dx.doi.org/10.3367/UFNe.0184.201402b.0137}{{\em Usp. Fiz. Nauk}
  {\bfseries 184} no.~2, (2014) 137--152},
  \href{http://arxiv.org/abs/1401.4024}{{\ttfamily arXiv:1401.4024 [hep-th]}}.

\bibitem{Libanov:2016kfc}
M.~Libanov, S.~Mironov, and V.~Rubakov, ``{Generalized Galileons: instabilities
  of bouncing and Genesis cosmologies and modified Genesis},''
  \href{http://dx.doi.org/10.1088/1475-7516/2016/08/037}{{\em JCAP} {\bfseries
  08} (2016) 037}, \href{http://arxiv.org/abs/1605.05992}{{\ttfamily
  arXiv:1605.05992 [hep-th]}}.

\bibitem{Kobayashi:2016xpl}
T.~Kobayashi, ``{Generic instabilities of nonsingular cosmologies in Horndeski
  theory: A no-go theorem},''
  \href{http://dx.doi.org/10.1103/PhysRevD.94.043511}{{\em Phys. Rev. D}
  {\bfseries 94} no.~4, (2016) 043511},
  \href{http://arxiv.org/abs/1606.05831}{{\ttfamily arXiv:1606.05831
  [hep-th]}}.

\bibitem{Ijjas:2016vtq}
A.~Ijjas and P.~J. Steinhardt, ``{Fully stable cosmological solutions with a
  non-singular classical bounce},''
  \href{http://dx.doi.org/10.1016/j.physletb.2016.11.047}{{\em Phys. Lett. B}
  {\bfseries 764} (2017) 289--294},
  \href{http://arxiv.org/abs/1609.01253}{{\ttfamily arXiv:1609.01253 [gr-qc]}}.

\bibitem{Dobre:2017pnt}
D.~A. Dobre, A.~V. Frolov, J.~T. G\'alvez~Ghersi, S.~Ramazanov, and A.~Vikman,
  ``{Unbraiding the Bounce: Superluminality around the Corner},''
  \href{http://dx.doi.org/10.1088/1475-7516/2018/03/020}{{\em JCAP} {\bfseries
  03} (2018) 020}, \href{http://arxiv.org/abs/1712.10272}{{\ttfamily
  arXiv:1712.10272 [gr-qc]}}.

\bibitem{Cai:2016thi}
Y.~Cai, Y.~Wan, H.-G. Li, T.~Qiu, and Y.-S. Piao, ``{The Effective Field Theory
  of nonsingular cosmology},''
  \href{http://dx.doi.org/10.1007/JHEP01(2017)090}{{\em JHEP} {\bfseries 01}
  (2017) 090}, \href{http://arxiv.org/abs/1610.03400}{{\ttfamily
  arXiv:1610.03400 [gr-qc]}}.

\bibitem{Creminelli:2016zwa}
P.~Creminelli, D.~Pirtskhalava, L.~Santoni, and E.~Trincherini, ``{Stability of
  Geodesically Complete Cosmologies},''
  \href{http://dx.doi.org/10.1088/1475-7516/2016/11/047}{{\em JCAP} {\bfseries
  11} (2016) 047}, \href{http://arxiv.org/abs/1610.04207}{{\ttfamily
  arXiv:1610.04207 [hep-th]}}.

\bibitem{Cai:2017tku}
Y.~Cai, H.-G. Li, T.~Qiu, and Y.-S. Piao, ``{The Effective Field Theory of
  nonsingular cosmology: II},''
  \href{http://dx.doi.org/10.1140/epjc/s10052-017-4938-y}{{\em Eur. Phys. J. C}
  {\bfseries 77} no.~6, (2017) 369},
  \href{http://arxiv.org/abs/1701.04330}{{\ttfamily arXiv:1701.04330 [gr-qc]}}.

\bibitem{Cai:2017dyi}
Y.~Cai and Y.-S. Piao, ``{A covariant Lagrangian for stable nonsingular
  bounce},'' \href{http://dx.doi.org/10.1007/JHEP09(2017)027}{{\em JHEP}
  {\bfseries 09} (2017) 027}, \href{http://arxiv.org/abs/1705.03401}{{\ttfamily
  arXiv:1705.03401 [gr-qc]}}.

\bibitem{Kolevatov:2017voe}
R.~Kolevatov, S.~Mironov, N.~Sukhov, and V.~Volkova, ``{Cosmological bounce and
  Genesis beyond Horndeski},''
  \href{http://dx.doi.org/10.1088/1475-7516/2017/08/038}{{\em JCAP} {\bfseries
  08} (2017) 038}, \href{http://arxiv.org/abs/1705.06626}{{\ttfamily
  arXiv:1705.06626 [hep-th]}}.

\bibitem{Ye:2019sth}
G.~Ye and Y.-S. Piao, ``{Bounce in general relativity and higher-order
  derivative operators},''
  \href{http://dx.doi.org/10.1103/PhysRevD.99.084019}{{\em Phys. Rev. D}
  {\bfseries 99} no.~8, (2019) 084019},
  \href{http://arxiv.org/abs/1901.08283}{{\ttfamily arXiv:1901.08283 [gr-qc]}}.

\bibitem{Arzoumanian:2020vkk}
{\bfseries NANOGrav} Collaboration, Z.~Arzoumanian {\em et~al.}, ``{The
  NANOGrav 12.5-year Data Set: Search For An Isotropic Stochastic
  Gravitational-Wave Background},''
  \href{http://arxiv.org/abs/2009.04496}{{\ttfamily arXiv:2009.04496
  [astro-ph.HE]}}.

\bibitem{Vagnozzi:2020gtf}
S.~Vagnozzi, ``{Implications of the NANOGrav results for inflation},''
  \href{http://arxiv.org/abs/2009.13432}{{\ttfamily arXiv:2009.13432
  [astro-ph.CO]}}.

\bibitem{Li:2020cjj}
H.-H. Li, G.~Ye, and Y.-S. Piao, ``{Is the NANOGrav signal a hint of dS decay
  during inflation?},'' \href{http://arxiv.org/abs/2009.14663}{{\ttfamily
  arXiv:2009.14663 [astro-ph.CO]}}.

\bibitem{Tahara:2020fmn}
H.~W. Tahara and T.~Kobayashi, ``{Nanohertz gravitational waves from NEC
  violation in the early universe},''
  \href{http://arxiv.org/abs/2011.01605}{{\ttfamily arXiv:2011.01605 [gr-qc]}}.

\bibitem{Kuroyanagi:2020sfw}
S.~Kuroyanagi, T.~Takahashi, and S.~Yokoyama, ``{Blue-tilted inflationary
  tensor spectrum and reheating in the light of NANOGrav results},''
  \href{http://arxiv.org/abs/2011.03323}{{\ttfamily arXiv:2011.03323
  [astro-ph.CO]}}.

\bibitem{Ade:2018gkx}
{\bfseries BICEP2, Keck Array} Collaboration, P.~Ade {\em et~al.}, ``{BICEP2 /
  Keck Array x: Constraints on Primordial Gravitational Waves using Planck,
  WMAP, and New BICEP2/Keck Observations through the 2015 Season},''
  \href{http://dx.doi.org/10.1103/PhysRevLett.121.221301}{{\em Phys. Rev.
  Lett.} {\bfseries 121} (2018) 221301},
  \href{http://arxiv.org/abs/1810.05216}{{\ttfamily arXiv:1810.05216
  [astro-ph.CO]}}.

\bibitem{Ashoorioon:2014nta}
A.~Ashoorioon, K.~Dimopoulos, M.~Sheikh-Jabbari, and G.~Shiu,
  ``{Non-Bunch\textendash{}Davis initial state reconciles chaotic models with
  BICEP and Planck},''
  \href{http://dx.doi.org/10.1016/j.physletb.2014.08.038}{{\em Phys. Lett. B}
  {\bfseries 737} (2014) 98--102},
  \href{http://arxiv.org/abs/1403.6099}{{\ttfamily arXiv:1403.6099 [hep-th]}}.

\bibitem{Piao:2003ty}
Y.-S. Piao and E.~Zhou, ``{Nearly scale invariant spectrum of adiabatic
  fluctuations may be from a very slowly expanding phase of the universe},''
  \href{http://dx.doi.org/10.1103/PhysRevD.68.083515}{{\em Phys. Rev. D}
  {\bfseries 68} (2003) 083515},
  \href{http://arxiv.org/abs/hep-th/0308080}{{\ttfamily arXiv:hep-th/0308080}}.

\bibitem{Piao:2004tq}
Y.-S. Piao and Y.-Z. Zhang, ``{Phantom inflation and primordial perturbation
  spectrum},'' \href{http://dx.doi.org/10.1103/PhysRevD.70.063513}{{\em Phys.
  Rev. D} {\bfseries 70} (2004) 063513},
  \href{http://arxiv.org/abs/astro-ph/0401231}{{\ttfamily
  arXiv:astro-ph/0401231}}.

\bibitem{Baldi:2005gk}
M.~Baldi, F.~Finelli, and S.~Matarrese, ``{Inflation with violation of the null
  energy condition},'' \href{http://dx.doi.org/10.1103/PhysRevD.72.083504}{{\em
  Phys. Rev. D} {\bfseries 72} (2005) 083504},
  \href{http://arxiv.org/abs/astro-ph/0505552}{{\ttfamily
  arXiv:astro-ph/0505552}}.

\bibitem{Piao:2006jz}
Y.-S. Piao, ``{Gravitational wave background from phantom superinflation},''
  \href{http://dx.doi.org/10.1103/PhysRevD.73.047302}{{\em Phys. Rev. D}
  {\bfseries 73} (2006) 047302},
  \href{http://arxiv.org/abs/gr-qc/0601115}{{\ttfamily arXiv:gr-qc/0601115}}.

\bibitem{Kobayashi:2010cm}
T.~Kobayashi, M.~Yamaguchi, and J.~Yokoyama, ``{G-inflation: Inflation driven
  by the Galileon field},''
  \href{http://dx.doi.org/10.1103/PhysRevLett.105.231302}{{\em Phys. Rev.
  Lett.} {\bfseries 105} (2010) 231302},
  \href{http://arxiv.org/abs/1008.0603}{{\ttfamily arXiv:1008.0603 [hep-th]}}.

\bibitem{Kobayashi:2011nu}
T.~Kobayashi, M.~Yamaguchi, and J.~Yokoyama, ``{Generalized G-inflation:
  Inflation with the most general second-order field equations},''
  \href{http://dx.doi.org/10.1143/PTP.126.511}{{\em Prog. Theor. Phys.}
  {\bfseries 126} (2011) 511--529},
  \href{http://arxiv.org/abs/1105.5723}{{\ttfamily arXiv:1105.5723 [hep-th]}}.

\bibitem{Creminelli:2006xe}
P.~Creminelli, M.~A. Luty, A.~Nicolis, and L.~Senatore, ``{Starting the
  Universe: Stable Violation of the Null Energy Condition and Non-standard
  Cosmologies},'' \href{http://dx.doi.org/10.1088/1126-6708/2006/12/080}{{\em
  JHEP} {\bfseries 12} (2006) 080},
  \href{http://arxiv.org/abs/hep-th/0606090}{{\ttfamily arXiv:hep-th/0606090}}.

\bibitem{Capurri:2020qgz}
G.~Capurri, N.~Bartolo, D.~Maino, and S.~Matarrese, ``{Let Effective Field
  Theory of Inflation flow: stochastic generation of models with red/blue
  tensor tilt},'' \href{http://dx.doi.org/10.1088/1475-7516/2020/11/037}{{\em
  JCAP} {\bfseries 11} (2020) 037},
  \href{http://arxiv.org/abs/2006.10781}{{\ttfamily arXiv:2006.10781
  [astro-ph.CO]}}.

\bibitem{Creminelli:2010ba}
P.~Creminelli, A.~Nicolis, and E.~Trincherini, ``{Galilean Genesis: An
  Alternative to inflation},''
  \href{http://dx.doi.org/10.1088/1475-7516/2010/11/021}{{\em JCAP} {\bfseries
  11} (2010) 021}, \href{http://arxiv.org/abs/1007.0027}{{\ttfamily
  arXiv:1007.0027 [hep-th]}}.

\bibitem{Liu:2011ns}
Z.-G. Liu, J.~Zhang, and Y.-S. Piao, ``{A Galileon Design of Slow Expansion},''
  \href{http://dx.doi.org/10.1103/PhysRevD.84.063508}{{\em Phys. Rev. D}
  {\bfseries 84} (2011) 063508},
  \href{http://arxiv.org/abs/1105.5713}{{\ttfamily arXiv:1105.5713
  [astro-ph.CO]}}.

\bibitem{Wang:2012bq}
Y.~Wang and R.~Brandenberger, ``{Scale-Invariant Fluctuations from Galilean
  Genesis},'' \href{http://dx.doi.org/10.1088/1475-7516/2012/10/021}{{\em JCAP}
  {\bfseries 10} (2012) 021}, \href{http://arxiv.org/abs/1206.4309}{{\ttfamily
  arXiv:1206.4309 [hep-th]}}.

\bibitem{Liu:2012ww}
Z.-G. Liu and Y.-S. Piao, ``{A Galileon Design of Slow Expansion: Emergent
  universe},'' \href{http://dx.doi.org/10.1016/j.physletb.2012.11.068}{{\em
  Phys. Lett. B} {\bfseries 718} (2013) 734--739},
  \href{http://arxiv.org/abs/1207.2568}{{\ttfamily arXiv:1207.2568 [gr-qc]}}.

\bibitem{Creminelli:2012my}
P.~Creminelli, K.~Hinterbichler, J.~Khoury, A.~Nicolis, and E.~Trincherini,
  ``{Subluminal Galilean Genesis},''
  \href{http://dx.doi.org/10.1007/JHEP02(2013)006}{{\em JHEP} {\bfseries 02}
  (2013) 006}, \href{http://arxiv.org/abs/1209.3768}{{\ttfamily arXiv:1209.3768
  [hep-th]}}.

\bibitem{Hinterbichler:2012fr}
K.~Hinterbichler, A.~Joyce, J.~Khoury, and G.~E. Miller, ``{DBI Realizations of
  the Pseudo-Conformal Universe and Galilean Genesis Scenarios},''
  \href{http://dx.doi.org/10.1088/1475-7516/2012/12/030}{{\em JCAP} {\bfseries
  12} (2012) 030}, \href{http://arxiv.org/abs/1209.5742}{{\ttfamily
  arXiv:1209.5742 [hep-th]}}.

\bibitem{Hinterbichler:2012yn}
K.~Hinterbichler, A.~Joyce, J.~Khoury, and G.~E. Miller, ``{Dirac-Born-Infeld
  Genesis: An Improved Violation of the Null Energy Condition},''
  \href{http://dx.doi.org/10.1103/PhysRevLett.110.241303}{{\em Phys. Rev.
  Lett.} {\bfseries 110} no.~24, (2013) 241303},
  \href{http://arxiv.org/abs/1212.3607}{{\ttfamily arXiv:1212.3607 [hep-th]}}.

\bibitem{Liu:2014tda}
Z.-G. Liu, H.~Li, and Y.-S. Piao, ``{Preinflationary genesis with CMB B-mode
  polarization},'' \href{http://dx.doi.org/10.1103/PhysRevD.90.083521}{{\em
  Phys. Rev. D} {\bfseries 90} no.~8, (2014) 083521},
  \href{http://arxiv.org/abs/1405.1188}{{\ttfamily arXiv:1405.1188
  [astro-ph.CO]}}.

\bibitem{Pirtskhalava:2014esa}
D.~Pirtskhalava, L.~Santoni, E.~Trincherini, and P.~Uttayarat, ``{Inflation
  from Minkowski Space},''
  \href{http://dx.doi.org/10.1007/JHEP12(2014)151}{{\em JHEP} {\bfseries 12}
  (2014) 151}, \href{http://arxiv.org/abs/1410.0882}{{\ttfamily arXiv:1410.0882
  [hep-th]}}.

\bibitem{Nishi:2015pta}
S.~Nishi and T.~Kobayashi, ``{Generalized Galilean Genesis},''
  \href{http://dx.doi.org/10.1088/1475-7516/2015/03/057}{{\em JCAP} {\bfseries
  03} (2015) 057}, \href{http://arxiv.org/abs/1501.02553}{{\ttfamily
  arXiv:1501.02553 [hep-th]}}.

\bibitem{Kobayashi:2015gga}
T.~Kobayashi, M.~Yamaguchi, and J.~Yokoyama, ``{Galilean Creation of the
  Inflationary Universe},''
  \href{http://dx.doi.org/10.1088/1475-7516/2015/07/017}{{\em JCAP} {\bfseries
  07} (2015) 017}, \href{http://arxiv.org/abs/1504.05710}{{\ttfamily
  arXiv:1504.05710 [hep-th]}}.

\bibitem{Cai:2016gjd}
Y.~Cai and Y.-S. Piao, ``{The slow expansion with nonminimal derivative
  coupling and its conformal dual},''
  \href{http://dx.doi.org/10.1007/JHEP03(2016)134}{{\em JHEP} {\bfseries 03}
  (2016) 134}, \href{http://arxiv.org/abs/1601.07031}{{\ttfamily
  arXiv:1601.07031 [hep-th]}}.

\bibitem{Nishi:2016ljg}
S.~Nishi and T.~Kobayashi, ``{Scale-invariant perturbations from
  null-energy-condition violation: A new variant of Galilean genesis},''
  \href{http://dx.doi.org/10.1103/PhysRevD.95.064001}{{\em Phys. Rev. D}
  {\bfseries 95} no.~6, (2017) 064001},
  \href{http://arxiv.org/abs/1611.01906}{{\ttfamily arXiv:1611.01906
  [hep-th]}}.

\bibitem{Mironov:2019qjt}
S.~Mironov, V.~Rubakov, and V.~Volkova, ``{Genesis with general relativity
  asymptotics in beyond Horndeski theory},''
  \href{http://dx.doi.org/10.1103/PhysRevD.100.083521}{{\em Phys. Rev. D}
  {\bfseries 100} no.~8, (2019) 083521},
  \href{http://arxiv.org/abs/1905.06249}{{\ttfamily arXiv:1905.06249
  [hep-th]}}.

\bibitem{Ageeva:2020gti}
Y.~Ageeva, O.~Evseev, O.~Melichev, and V.~Rubakov, ``{Toward evading the strong
  coupling problem in Horndeski genesis},''
  \href{http://dx.doi.org/10.1103/PhysRevD.102.023519}{{\em Phys. Rev. D}
  {\bfseries 102} no.~2, (2020) 023519},
  \href{http://arxiv.org/abs/2003.01202}{{\ttfamily arXiv:2003.01202
  [hep-th]}}.

\bibitem{Ilyas:2020zcb}
A.~Ilyas, M.~Zhu, Y.~Zheng, and Y.-F. Cai, ``{Emergent Universe and Genesis
  from the DHOST Cosmology},''
  \href{http://arxiv.org/abs/2009.10351}{{\ttfamily arXiv:2009.10351 [gr-qc]}}.

\bibitem{Langlois:2018dxi}
D.~Langlois, ``{Dark energy and modified gravity in degenerate higher-order
  scalar\textendash{}tensor (DHOST) theories: A review},''
  \href{http://dx.doi.org/10.1142/S0218271819420069}{{\em Int. J. Mod. Phys. D}
  {\bfseries 28} no.~05, (2019) 1942006},
  \href{http://arxiv.org/abs/1811.06271}{{\ttfamily arXiv:1811.06271 [gr-qc]}}.

\bibitem{Kobayashi:2019hrl}
T.~Kobayashi, ``{Horndeski theory and beyond: a review},''
  \href{http://dx.doi.org/10.1088/1361-6633/ab2429}{{\em Rept. Prog. Phys.}
  {\bfseries 82} no.~8, (2019) 086901},
  \href{http://arxiv.org/abs/1901.07183}{{\ttfamily arXiv:1901.07183 [gr-qc]}}.

\bibitem{Liu:2013xt}
Z.-G. Liu and Y.-S. Piao, ``{Galilean Islands in Eternally Inflating
  Background},'' \href{http://dx.doi.org/10.1103/PhysRevD.88.043520}{{\em Phys.
  Rev. D} {\bfseries 88} (2013) 043520},
  \href{http://arxiv.org/abs/1301.6833}{{\ttfamily arXiv:1301.6833 [gr-qc]}}.

\bibitem{Alberte:2016izw}
L.~Alberte, P.~Creminelli, A.~Khmelnitsky, D.~Pirtskhalava, and E.~Trincherini,
  ``{Relaxing the Cosmological Constant: a Proof of Concept},''
  \href{http://dx.doi.org/10.1007/JHEP12(2016)022}{{\em JHEP} {\bfseries 12}
  (2016) 022}, \href{http://arxiv.org/abs/1608.05715}{{\ttfamily
  arXiv:1608.05715 [hep-th]}}.

\bibitem{Rubakov:2013kaa}
V.~Rubakov, ``{Consistent NEC-violation: towards creating a universe in the
  laboratory},'' \href{http://dx.doi.org/10.1103/PhysRevD.88.044015}{{\em Phys.
  Rev. D} {\bfseries 88} (2013) 044015},
  \href{http://arxiv.org/abs/1305.2614}{{\ttfamily arXiv:1305.2614 [hep-th]}}.

\bibitem{Elder:2013gya}
B.~Elder, A.~Joyce, and J.~Khoury, ``{From Satisfying to Violating the Null
  Energy Condition},'' \href{http://dx.doi.org/10.1103/PhysRevD.89.044027}{{\em
  Phys. Rev. D} {\bfseries 89} no.~4, (2014) 044027},
  \href{http://arxiv.org/abs/1311.5889}{{\ttfamily arXiv:1311.5889 [hep-th]}}.

\bibitem{Buchbinder:2007ad}
E.~I. Buchbinder, J.~Khoury, and B.~A. Ovrut, ``{New Ekpyrotic cosmology},''
  \href{http://dx.doi.org/10.1103/PhysRevD.76.123503}{{\em Phys. Rev. D}
  {\bfseries 76} (2007) 123503},
  \href{http://arxiv.org/abs/hep-th/0702154}{{\ttfamily arXiv:hep-th/0702154}}.

\bibitem{Koehn:2015vvy}
M.~Koehn, J.-L. Lehners, and B.~Ovrut, ``{Nonsingular bouncing cosmology:
  Consistency of the effective description},''
  \href{http://dx.doi.org/10.1103/PhysRevD.93.103501}{{\em Phys. Rev. D}
  {\bfseries 93} no.~10, (2016) 103501},
  \href{http://arxiv.org/abs/1512.03807}{{\ttfamily arXiv:1512.03807
  [hep-th]}}.

\bibitem{Cai:2015yza}
Y.~Cai, Y.-T. Wang, and Y.-S. Piao, ``{Is there an effect of a nontrivial $c_T$
  during inflation?},''
  \href{http://dx.doi.org/10.1103/PhysRevD.93.063005}{{\em Phys. Rev. D}
  {\bfseries 93} no.~6, (2016) 063005},
  \href{http://arxiv.org/abs/1510.08716}{{\ttfamily arXiv:1510.08716
  [astro-ph.CO]}}.

\bibitem{Cai:2016ldn}
Y.~Cai, Y.-T. Wang, and Y.-S. Piao, ``{Propagating speed of primordial
  gravitational waves and inflation},''
  \href{http://dx.doi.org/10.1103/PhysRevD.94.043002}{{\em Phys. Rev. D}
  {\bfseries 94} no.~4, (2016) 043002},
  \href{http://arxiv.org/abs/1602.05431}{{\ttfamily arXiv:1602.05431
  [astro-ph.CO]}}.

\bibitem{Giovannini:2015kfa}
M.~Giovannini, ``{The refractive index of relic gravitons},''
  \href{http://dx.doi.org/10.1088/0264-9381/33/12/125002}{{\em Class. Quant.
  Grav.} {\bfseries 33} no.~12, (2016) 125002},
  \href{http://arxiv.org/abs/1507.03456}{{\ttfamily arXiv:1507.03456
  [astro-ph.CO]}}.

\bibitem{Giovannini:2018zbf}
M.~Giovannini, ``{The propagating speed of relic gravitational waves and their
  refractive index during inflation},''
  \href{http://dx.doi.org/10.1140/epjc/s10052-018-5931-9}{{\em Eur. Phys. J. C}
  {\bfseries 78} no.~6, (2018) 442},
  \href{http://arxiv.org/abs/1803.05203}{{\ttfamily arXiv:1803.05203 [gr-qc]}}.

\bibitem{Giare:2020vss}
W.~Giar\`e and F.~Renzi, ``{Propagating speed of primordial gravitational
  waves},'' \href{http://dx.doi.org/10.1103/PhysRevD.102.083530}{{\em Phys.
  Rev. D} {\bfseries 102} no.~8, (2020) 083530},
  \href{http://arxiv.org/abs/2007.04256}{{\ttfamily arXiv:2007.04256
  [astro-ph.CO]}}.

\bibitem{Giare:2020plo}
W.~Giar\`e, F.~Renzi, and A.~Melchiorri, ``{Higher-Curvature Corrections and
  Tensor Modes},'' \href{http://arxiv.org/abs/2012.00527}{{\ttfamily
  arXiv:2012.00527 [astro-ph.CO]}}.

\bibitem{Cai:2015nya}
Y.~Cai, Y.-T. Wang, and Y.-S. Piao, ``{Preinflationary primordial
  perturbations},'' \href{http://dx.doi.org/10.1103/PhysRevD.92.023518}{{\em
  Phys. Rev. D} {\bfseries 92} no.~2, (2015) 023518},
  \href{http://arxiv.org/abs/1501.01730}{{\ttfamily arXiv:1501.01730
  [astro-ph.CO]}}.

\bibitem{Cai:2019hge}
Y.~Cai and Y.-S. Piao, ``{Pre-inflation and trans-Planckian censorship},''
  \href{http://dx.doi.org/10.1007/s11433-020-1573-5}{{\em Sci. China Phys.
  Mech. Astron.} {\bfseries 63} no.~11, (2020) 110411},
  \href{http://arxiv.org/abs/1909.12719}{{\ttfamily arXiv:1909.12719 [gr-qc]}}.

\bibitem{Piao:2003zm}
Y.-S. Piao, B.~Feng, and X.-m. Zhang, ``{Suppressing CMB quadrupole with a
  bounce from contracting phase to inflation},''
  \href{http://dx.doi.org/10.1103/PhysRevD.69.103520}{{\em Phys. Rev. D}
  {\bfseries 69} (2004) 103520},
  \href{http://arxiv.org/abs/hep-th/0310206}{{\ttfamily arXiv:hep-th/0310206}}.

\bibitem{Piao:2005ag}
Y.-S. Piao, ``{A Possible explanation to low CMB quadrupole},''
  \href{http://dx.doi.org/10.1103/PhysRevD.71.087301}{{\em Phys. Rev. D}
  {\bfseries 71} (2005) 087301},
  \href{http://arxiv.org/abs/astro-ph/0502343}{{\ttfamily
  arXiv:astro-ph/0502343}}.

\bibitem{Liu:2013kea}
Z.-G. Liu, Z.-K. Guo, and Y.-S. Piao, ``{Obtaining the CMB anomalies with a
  bounce from the contracting phase to inflation},''
  \href{http://dx.doi.org/10.1103/PhysRevD.88.063539}{{\em Phys. Rev. D}
  {\bfseries 88} (2013) 063539},
  \href{http://arxiv.org/abs/1304.6527}{{\ttfamily arXiv:1304.6527
  [astro-ph.CO]}}.

\bibitem{Qiu:2015nha}
T.~Qiu and Y.-T. Wang, ``{G-Bounce Inflation: Towards Nonsingular Inflation
  Cosmology with Galileon Field},''
  \href{http://dx.doi.org/10.1007/JHEP04(2015)130}{{\em JHEP} {\bfseries 04}
  (2015) 130}, \href{http://arxiv.org/abs/1501.03568}{{\ttfamily
  arXiv:1501.03568 [astro-ph.CO]}}.

\bibitem{Cai:2017pga}
Y.~Cai, Y.-T. Wang, J.-Y. Zhao, and Y.-S. Piao, ``{Primordial perturbations
  with pre-inflationary bounce},''
  \href{http://dx.doi.org/10.1103/PhysRevD.97.103535}{{\em Phys. Rev. D}
  {\bfseries 97} no.~10, (2018) 103535},
  \href{http://arxiv.org/abs/1709.07464}{{\ttfamily arXiv:1709.07464
  [astro-ph.CO]}}.

\bibitem{Ellis:2020ena}
J.~Ellis and M.~Lewicki, ``{Cosmic String Interpretation of NANOGrav Pulsar
  Timing Data},'' \href{http://arxiv.org/abs/2009.06555}{{\ttfamily
  arXiv:2009.06555 [astro-ph.CO]}}.

\bibitem{Blasi:2020mfx}
S.~Blasi, V.~Brdar, and K.~Schmitz, ``{Has NANOGrav found first evidence for
  cosmic strings?},'' \href{http://arxiv.org/abs/2009.06607}{{\ttfamily
  arXiv:2009.06607 [astro-ph.CO]}}.

\bibitem{DeLuca:2020agl}
V.~De~Luca, G.~Franciolini, and A.~Riotto, ``{NANOGrav Hints to Primordial
  Black Holes as Dark Matter},''
  \href{http://arxiv.org/abs/2009.08268}{{\ttfamily arXiv:2009.08268
  [astro-ph.CO]}}.

\bibitem{Buchmuller:2020lbh}
W.~Buchmuller, V.~Domcke, and K.~Schmitz, ``{From NANOGrav to LIGO with
  metastable cosmic strings},''
  \href{http://dx.doi.org/10.1016/j.physletb.2020.135914}{{\em Phys. Lett. B}
  {\bfseries 811} (2020) 135914},
  \href{http://arxiv.org/abs/2009.10649}{{\ttfamily arXiv:2009.10649
  [astro-ph.CO]}}.

\bibitem{Addazi:2020zcj}
A.~Addazi, Y.-F. Cai, Q.~Gan, A.~Marciano, and K.~Zeng, ``{NANOGrav results and
  Dark First Order Phase Transitions},''
  \href{http://arxiv.org/abs/2009.10327}{{\ttfamily arXiv:2009.10327
  [hep-ph]}}.

\bibitem{Kohri:2020qqd}
K.~Kohri and T.~Terada, ``{Solar-Mass Primordial Black Holes Explain NANOGrav
  Hint of Gravitational Waves},''
  \href{http://arxiv.org/abs/2009.11853}{{\ttfamily arXiv:2009.11853
  [astro-ph.CO]}}.

\bibitem{Ratzinger:2020koh}
W.~Ratzinger and P.~Schwaller, ``{Whispers from the dark side: Confronting
  light new physics with NANOGrav data},''
  \href{http://arxiv.org/abs/2009.11875}{{\ttfamily arXiv:2009.11875
  [astro-ph.CO]}}.

\bibitem{Samanta:2020cdk}
R.~Samanta and S.~Datta, ``{Gravitational wave complementarity and impact of
  NANOGrav data on gravitational leptogenesis: cosmic strings},''
  \href{http://arxiv.org/abs/2009.13452}{{\ttfamily arXiv:2009.13452
  [hep-ph]}}.

\bibitem{Bian:2020bps}
L.~Bian, J.~Liu, and R.~Zhou, ``{NanoGrav 12.5-yr data and different stochastic
  Gravitational wave background sources},''
  \href{http://arxiv.org/abs/2009.13893}{{\ttfamily arXiv:2009.13893
  [astro-ph.CO]}}.

\bibitem{Neronov:2020qrl}
A.~Neronov, A.~Roper~Pol, C.~Caprini, and D.~Semikoz, ``{NANOGrav signal from
  MHD turbulence at QCD phase transition in the early universe},''
  \href{http://arxiv.org/abs/2009.14174}{{\ttfamily arXiv:2009.14174
  [astro-ph.CO]}}.

\bibitem{Sugiyama:2020roc}
S.~Sugiyama, V.~Takhistov, E.~Vitagliano, A.~Kusenko, M.~Sasaki, and M.~Takada,
  ``{Testing Stochastic Gravitational Wave Signals from Primordial Black Holes
  with Optical Telescopes},'' \href{http://arxiv.org/abs/2010.02189}{{\ttfamily
  arXiv:2010.02189 [astro-ph.CO]}}.

\bibitem{Liu:2020mru}
J.~Liu, R.-G. Cai, and Z.-K. Guo, ``{Large anisotropies of the stochastic
  gravitational wave background from cosmic domain walls},''
  \href{http://arxiv.org/abs/2010.03225}{{\ttfamily arXiv:2010.03225
  [astro-ph.CO]}}.

\bibitem{Domenech:2020ers}
G.~Dom\`enech and S.~Pi, ``{NANOGrav Hints on Planet-Mass Primordial Black
  Holes},'' \href{http://arxiv.org/abs/2010.03976}{{\ttfamily arXiv:2010.03976
  [astro-ph.CO]}}.

\bibitem{Bhattacharya:2020lhc}
S.~Bhattacharya, S.~Mohanty, and P.~Parashari, ``{Implications of the NANOGrav
  result on primordial gravitational waves in nonstandard cosmologies},''
  \href{http://arxiv.org/abs/2010.05071}{{\ttfamily arXiv:2010.05071
  [astro-ph.CO]}}.

\bibitem{Wong:2020yig}
K.~Wong, G.~Franciolini, V.~De~Luca, V.~Baibhav, E.~Berti, P.~Pani, and
  A.~Riotto, ``{Constraining the primordial black hole scenario with Bayesian
  inference and machine learning: the GWTC-2 gravitational wave catalog},''
  \href{http://arxiv.org/abs/2011.01865}{{\ttfamily arXiv:2011.01865 [gr-qc]}}.

\bibitem{Turner:1993vb}
M.~S. Turner, M.~J. White, and J.~E. Lidsey, ``{Tensor perturbations in
  inflationary models as a probe of cosmology},''
  \href{http://dx.doi.org/10.1103/PhysRevD.48.4613}{{\em Phys. Rev. D}
  {\bfseries 48} (1993) 4613--4622},
  \href{http://arxiv.org/abs/astro-ph/9306029}{{\ttfamily
  arXiv:astro-ph/9306029}}.

\bibitem{Boyle:2005se}
L.~A. Boyle and P.~J. Steinhardt, ``{Probing the early universe with
  inflationary gravitational waves},''
  \href{http://dx.doi.org/10.1103/PhysRevD.77.063504}{{\em Phys. Rev. D}
  {\bfseries 77} (2008) 063504},
  \href{http://arxiv.org/abs/astro-ph/0512014}{{\ttfamily
  arXiv:astro-ph/0512014}}.

\bibitem{Zhao:2006mm}
W.~Zhao and Y.~Zhang, ``{Relic gravitational waves and their detection},''
  \href{http://dx.doi.org/10.1103/PhysRevD.74.043503}{{\em Phys. Rev. D}
  {\bfseries 74} (2006) 043503},
  \href{http://arxiv.org/abs/astro-ph/0604458}{{\ttfamily
  arXiv:astro-ph/0604458}}.

\bibitem{Kuroyanagi:2014nba}
S.~Kuroyanagi, T.~Takahashi, and S.~Yokoyama, ``{Blue-tilted Tensor Spectrum
  and Thermal History of the Universe},''
  \href{http://dx.doi.org/10.1088/1475-7516/2015/02/003}{{\em JCAP} {\bfseries
  02} (2015) 003}, \href{http://arxiv.org/abs/1407.4785}{{\ttfamily
  arXiv:1407.4785 [astro-ph.CO]}}.

\bibitem{Liu:2015psa}
X.-J. Liu, W.~Zhao, Y.~Zhang, and Z.-H. Zhu, ``{Detecting Relic Gravitational
  Waves by Pulsar Timing Arrays: Effects of Cosmic Phase Transitions and
  Relativistic Free-Streaming Gases},''
  \href{http://dx.doi.org/10.1103/PhysRevD.93.024031}{{\em Phys. Rev. D}
  {\bfseries 93} no.~2, (2016) 024031},
  \href{http://arxiv.org/abs/1509.03524}{{\ttfamily arXiv:1509.03524
  [astro-ph.CO]}}.

\bibitem{Adams:1997de}
J.~A. Adams, G.~G. Ross, and S.~Sarkar, ``{Multiple inflation},''
  \href{http://dx.doi.org/10.1016/S0550-3213(97)00431-8}{{\em Nucl. Phys. B}
  {\bfseries 503} (1997) 405--425},
  \href{http://arxiv.org/abs/hep-ph/9704286}{{\ttfamily arXiv:hep-ph/9704286}}.

\bibitem{Burgess:2005sb}
C.~Burgess, R.~Easther, A.~Mazumdar, D.~F. Mota, and T.~Multamaki, ``{Multiple
  inflation, cosmic string networks and the string landscape},''
  \href{http://dx.doi.org/10.1088/1126-6708/2005/05/067}{{\em JHEP} {\bfseries
  05} (2005) 067}, \href{http://arxiv.org/abs/hep-th/0501125}{{\ttfamily
  arXiv:hep-th/0501125}}.

\bibitem{Ashoorioon:2006wc}
A.~Ashoorioon and A.~Krause, ``{Power Spectrum and Signatures for Cascade
  Inflation},'' \href{http://arxiv.org/abs/hep-th/0607001}{{\ttfamily
  arXiv:hep-th/0607001}}.

\bibitem{Ashoorioon:2008qr}
A.~Ashoorioon, A.~Krause, and K.~Turzynski, ``{Energy Transfer in Multi Field
  Inflation and Cosmological Perturbations},''
  \href{http://dx.doi.org/10.1088/1475-7516/2009/02/014}{{\em JCAP} {\bfseries
  02} (2009) 014}, \href{http://arxiv.org/abs/0810.4660}{{\ttfamily
  arXiv:0810.4660 [hep-th]}}.

\bibitem{Liu:2009pk}
Y.~Liu, Y.-S. Piao, and Z.-G. Si, ``{'Old' Locked Inflation},''
  \href{http://dx.doi.org/10.1088/1475-7516/2009/05/008}{{\em JCAP} {\bfseries
  05} (2009) 008}, \href{http://arxiv.org/abs/0901.2058}{{\ttfamily
  arXiv:0901.2058 [hep-th]}}.

\bibitem{DAmico:2020euu}
G.~D'Amico and N.~Kaloper, ``{Rollercoaster Cosmology},''
  \href{http://arxiv.org/abs/2011.09489}{{\ttfamily arXiv:2011.09489
  [hep-th]}}.

\bibitem{Obied:2018sgi}
G.~Obied, H.~Ooguri, L.~Spodyneiko, and C.~Vafa, ``{De Sitter Space and the
  Swampland},'' \href{http://arxiv.org/abs/1806.08362}{{\ttfamily
  arXiv:1806.08362 [hep-th]}}.

\bibitem{Bedroya:2019snp}
A.~Bedroya and C.~Vafa, ``{Trans-Planckian Censorship and the Swampland},''
  \href{http://dx.doi.org/10.1007/JHEP09(2020)123}{{\em JHEP} {\bfseries 09}
  (2020) 123}, \href{http://arxiv.org/abs/1909.11063}{{\ttfamily
  arXiv:1909.11063 [hep-th]}}.

\bibitem{Li:2019ipk}
H.-H. Li, G.~Ye, Y.~Cai, and Y.-S. Piao, ``{Trans-Planckian censorship of
  multistage inflation and dark energy},''
  \href{http://dx.doi.org/10.1103/PhysRevD.101.063527}{{\em Phys. Rev. D}
  {\bfseries 101} no.~6, (2020) 063527},
  \href{http://arxiv.org/abs/1911.06148}{{\ttfamily arXiv:1911.06148 [gr-qc]}}.

\bibitem{Berera:2019zdd}
A.~Berera and J.~R. Calder\'on, ``{Trans-Planckian censorship and other
  swampland bothers addressed in warm inflation},''
  \href{http://dx.doi.org/10.1103/PhysRevD.100.123530}{{\em Phys. Rev. D}
  {\bfseries 100} no.~12, (2019) 123530},
  \href{http://arxiv.org/abs/1910.10516}{{\ttfamily arXiv:1910.10516
  [hep-ph]}}.

\bibitem{Dhuria:2019oyf}
M.~Dhuria and G.~Goswami, ``{Trans-Planckian censorship conjecture and
  nonthermal post-inflationary history},''
  \href{http://dx.doi.org/10.1103/PhysRevD.100.123518}{{\em Phys. Rev. D}
  {\bfseries 100} no.~12, (2019) 123518},
  \href{http://arxiv.org/abs/1910.06233}{{\ttfamily arXiv:1910.06233
  [astro-ph.CO]}}.

\bibitem{Torabian:2019zms}
M.~Torabian, ``{Non-Standard Cosmological Models and The Trans-Planckian
  Censorship Conjecture},''
  \href{http://dx.doi.org/10.1002/prop.201900092}{{\em Fortsch. Phys.}
  {\bfseries 68} no.~2, (2020) 1900092},
  \href{http://arxiv.org/abs/1910.06867}{{\ttfamily arXiv:1910.06867
  [hep-th]}}.

\bibitem{Cai:2017dxl}
Y.~Cai and Y.-S. Piao, ``{Higher order derivative coupling to gravity and its
  cosmological implications},''
  \href{http://dx.doi.org/10.1103/PhysRevD.96.124028}{{\em Phys. Rev. D}
  {\bfseries 96} no.~12, (2017) 124028},
  \href{http://arxiv.org/abs/1707.01017}{{\ttfamily arXiv:1707.01017 [gr-qc]}}.

\bibitem{Ye:2019frg}
G.~Ye and Y.-S. Piao, ``{Implication of GW170817 for cosmological bounces},''
  \href{http://dx.doi.org/10.1088/0253-6102/71/4/427}{{\em Commun. Theor.
  Phys.} {\bfseries 71} no.~4, (2019) 427},
  \href{http://arxiv.org/abs/1901.02202}{{\ttfamily arXiv:1901.02202 [gr-qc]}}.

\end{thebibliography}\endgroup

\end{document}